\pdfoutput=1
\documentclass[aps,reprint,groupedaddress,a4paper,twoside,twocolumn]{revtex4-1}
\usepackage[paperwidth=210mm,paperheight=297mm,centering,hmargin=2cm,vmargin=2.5cm]{geometry}
\usepackage{graphicx}
\usepackage{color}
\graphicspath{{img/}}
\usepackage{url}
\newcommand{\com}[1]{{#1}}
\usepackage{gensymb}
\usepackage{bm}
\begin{document}

\title{\Large Coarse-Grained Simulation of DNA using LAMMPS\\
\large An implementation of the oxDNA model and its applications\\}
\author{Oliver Henrich$^*$}
\affiliation{Department of Physics, SUPA, University of Strathclyde, Glasgow G4 0NG, Scotland, UK}
\email{Corresponding email address: oliver.henrich@strath.ac.uk}
\author{Yair Augusto Guti\'errez Fosado}
\affiliation{School of Physics and Astronomy, University of Edinburgh, Edinburgh EH9 3FD, Scotland, UK}
\author{Tine Curk}
\affiliation{CAS Key Laboratory of Soft Matter Physics, Beijing National Laboratory for Condensed Matter Physics, Institute of Physics, Chinese Academy of Sciences, Beijing 100190, China \&\\
Department of Chemistry, University of Cambridge, Cambridge CB2 1EW, UK}
\author{Thomas E. Ouldridge}
\affiliation{Department of Bioengineering \& Centre of Synthetic Biology, Imperial College London, London SW7 2AZ, UK}

%}                     % Do not remove
%\affiliation{
%Department of Physics, SUPA, University of Strathclyde, Glasgow G4 0NG, Scotland, UK \and
%School of Physics and Astronomy, University of Edinburgh, Edinburgh EH9 3FD, Scotland, UK \and
%CAS Key Laboratory of Soft Matter Physics, Beijing National Laboratory for Condensed Matter Physics, Institute of Physics, Chinese Academy of Sciences, Beijing 100190, China \and
%Department of Chemistry, University of Cambridge, Cambridge CB2 1EW, UK \and
%Department of Bioengineering, Imperial College London, London SW7 2AZ, UK\\
%{*} Email: oliver.henrich@strath.ac.uk
%}
%

%\date{\today}
\begin{abstract}
During the last decade coarse-grained nucleotide models have emerged that allow us to study DNA and RNA on unprecedented time and length scales. Among them is oxDNA, a coarse-grained, sequence-specific model that captures the hybridisation transition of DNA and many structural properties of single- and double-stranded DNA. oxDNA was previously only available as standalone software, but has now been implemented into the popular LAMMPS molecular dynamics code. This article describes the new implementation and analyses its parallel performance. Practical applications are presented that focus on single-stranded DNA, an area of research which has been so far under-investigated. The LAMMPS implementation of oxDNA lowers the entry barrier for using the oxDNA model significantly, facilitates future code development and interfacing with existing LAMMPS functionality as well as other coarse-grained and atomistic DNA models. %
\end{abstract}

 % end of PACS codes

\maketitle
%

%%%%%%%%%%%%%%%%%%%%%%%%%%%%%%%%%%%%%%%%%%%%%%%%%
% Introduction
%%%%%%%%%%%%%%%%%%%%%%%%%%%%%%%%%%%%%%%%%%%%%%%%%

\section{Introduction}

DNA is one of the most important bio-polymers, as its sequence encodes the genetic instructions needed in the development and functioning of many living organisms. While we know now the sequence of many genomes, we still know little as to how DNA is organised in 3D inside a living cell, and of how gene regulation and DNA function are coupled to this structure.
The complexity of the DNA molecule can be brought to mind by highlighting a few of its quantitative aspects. The entire DNA within a single human cell is about 2\,m long, but only 2\,nm wide and organised at different hierarchical levels. If compressed into a spherical ball, this ball would have a diameter of about 2\,$\mu$m \cite{Calladine}.

Computational modelling of DNA appears as the only avenue to understanding its intricacies in sufficient detail and has been an important field in biophysics for decades. Traditionally, most of the available simulation techniques have worked at the atomistic level of detail \cite{Laughton:2011}. Existing atomistic force fields can capture fast conformational fluctuations and protein-DNA binding, but cannot deliver the necessary temporal and spatial resolution to describe phenomena that occur on larger time and length scales as they are often limited to a few hundred base pairs and (at most) microsecond time scales. 
Recent years have therefore witnessed a rapid increase of a new research effort at a different, coarse-grained level  \cite{Potoyan:2013}. Coarse-grained (CG) models of DNA 
can provide significant computational and conceptual advantages over atomistic models leading often to three or more orders of magnitude greater efficiency. The challenge consists in retaining the right degrees of freedom so that the CG model reproduces relevant emergent structural features and thermodynamic properties of DNA. CG modelling of DNA is not only an efficient alternative to atomistic approaches. It is indispensable for the modelling of DNA on timescales in the millisecond range and beyond, or when long DNA strands of tens of thousands of base pairs or more have to be considered, e.g. to study the dynamics of DNA supercoiling (i.e. the local over- or under-twisting of the double helix, which is also important for gene expression in bacteria), of genomic DNA loops and of chromatin or chromosome fragments.

A small number of very promising CG DNA models have emerged to date. Conceptually they can be categorised into top-down approaches, which use empirical interactions that are parameterised to match experimental observables, or bottom-up approaches, which eliminate dispensable degrees of freedom systematically starting from atomistic force fields. They may also target different applications depending on their capabilities, such as single versus double stranded DNA (ssDNA and dsDNA), or nanotechnological versus biological applications. We refer to \cite{Dans:2016} for a comprehensive overview of the capabilities of individual models and recent activities in this field.

From a software point of view these models are often based on standalone software \cite{Maffeo:2014,Korolev:2014,Maciejczyk:2014}, which has a somewhat limiting effect on uptake and user communities growth. Others models use popular MD-codes as computational platforms, such as GROMACS \cite{GROMACS} in case of the SIRAH \cite{Machado:2016} and the MARTINI force field \cite{Uusitalo:2015}, or NAMD \cite{NAMD,Markegard:2015}. Another suitable platform for CG simulation of DNA has emerged in form of the powerful Large-Scale Atomic/Mo\-lecu\-lar Massively Parallel Simulator (LAMMPS) for molecular dynamics \cite{lammps}, including the widely used 3SPN.2 model \cite{Hinckley:2013,Hinckley:2015} and others that target even larger length scales \cite{Brackley:2014,Fosado:2016}. 

This article reports the latest effort of implementing the popular oxDNA model \cite{Ouldridge:2011,tom-thesis} into the LAMMPS code. Until recently this model was only available as bespoke and standalone software \cite{oxdna}. Through the efficient parallelisation of LAMMPS it is now possible to run oxDNA in parallel on multi-core CPU-architectures, extending its capabilities to unprecedented time and length scales. The largest system that could be studied by oxDNA was previously limited by the size of system that can be fitted onto a single GPU.

This paper is organised as follows: In Section \ref{model} we briefly introduce the details of the oxDNA and oxDNA2 models. Section \ref{implementation} explains how the LAMMPS implementation of the oxDNA models can be invoked and provides further information on the code distribution and documentation. In Section \ref{integrator} we describe the LAMMPS implementation of novel Langevin-type rigid body integrators which feature improved stability and accuracy. Section \ref{performance} gives details of the scaling performance of parallel implementation. Section \ref{applications} presents results on the behaviour of single-stranded DNA, an area of DNA research which so far has not been intensively investigated. One application is concerned with lambda-DNA of a bacteriophage, whereas the other application involves a plasmid cloning vector pUC19. In Section \ref{conclusions} we summarise this work.

%%%%%%%%%%%%%%%%%%%%%%%%%%%%%%%%%%%%%%%%%%%%%%%%%
% Model
%%%%%%%%%%%%%%%%%%%%%%%%%%%%%%%%%%%%%%%%%%%%%%%%%

\section{The oxDNA Model}\label{model}

\begin{figure}[htpb]
\begin{center}
\includegraphics[width=0.5\textwidth]{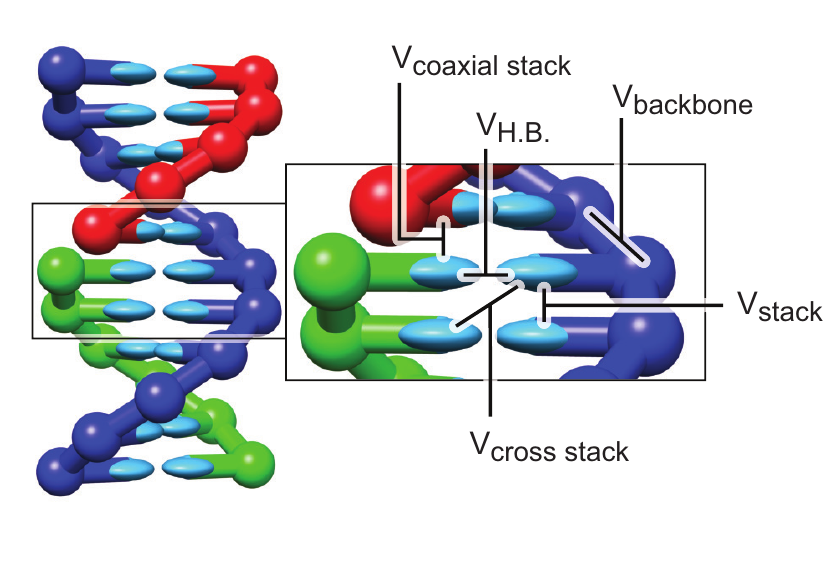}
\caption{\label{oxdna_detail} Overview of bonded and pair interactions in oxDNA: phosphate backbone connectivity and excluded volume, hydrogen-bonding, stacking, cross-stacking and coaxial stacking interaction. The oxDNA2 model contains an additional implicit electrostatic interaction in form of a Debye-H\"uckel potential. 
Reprinted from \cite{Ouldridge:2013} with permission from ACS Nano. Copyright (2013) American Chemical Society.}
\end{center}
\end{figure}

The oxDNA model consists of rigid nucleotides with three interaction sites for the effective interactions between the nucleotides. These pairwise-additive forces arise 
due to the excluded volume, the connectivity of the phosphate backbone, the stacking, cross-stacking and coaxial stacking as a 
consequence of the hydrophobicity of the bases, as well as hydrogen bonding between complementary base pairs. 
Fig. \ref{oxdna_detail} illustrates these interactions schematically for the original version of the model, to 
which we refer as oxDNA \cite{tom-thesis}. 
In this version all three interaction sites are co-linear. The hydrogen bonding/excluded volume site and the stacking site are separated 
from the backbone/electrostatic interaction site by $0.74$ length units ($6.3$\,\AA) and $0.8$ length units ($6.8$\,\AA), respectively. 
The orientation of the bases is specified by a base normal vector, which defines the notional plane of the base and the vector between the interaction sites. 
Together with the relative distance vectors between the interaction sites, the base vector and base normal vector are used to modulate 
the stacking, cross-stacking, coaxial stacking and hydrogen bonding interaction between two consecutive nucleotides.
  
The simplest interaction is the backbone connectivity, which is modelled with FENE (finitely extensible nonlinear elastic) springs acting 
between the backbone interaction sites. The excluded volume interaction is modelled with truncated and smoothed Lennard-Jones potentials between backbone sites, base sites and between the backbone and base sites. The hydrogen bonding interaction consists of smoothed, truncated and modulated Morse potentials between the hydrogen bonding site.
The stacking interaction falls into three individual sub-interactions: the stacking interaction between consecutive nucleotides on the same strand as well as cross-stacking and coaxial stacking between any nucleotide in the appropriate relative position. 
It is worth emphasising that the duplex structure is not specified or imposed in any other way, but emerges naturally through this choice of interactions and their parameterisation. This is another strength of the oxDNA model and permits an accurate description of both ssDNA and dsDNA. 
The stacking interactions are modelled with a combination of smoothed, truncated and modulated Morse, harmonic angle and harmonic distance potentials. 
All interactions have been parameterised to match key thermodynamic properties of ssDNA and dsDNA such as the longitudinal and torsional persistence 
length or the melting temperature of the duplex \cite{Ouldridge:2011,Holbrook:1999,SantaLucia:2004}.

A short schematic overview of various interactions involved in the definition of oxDNA model is given in Fig. \ref{oxdna_detail}. More details can be found in the original publications \cite{Ouldridge:2011,tom-thesis}.

\begin{figure}[htpb]
\begin{center}
\includegraphics[width=0.5\textwidth]{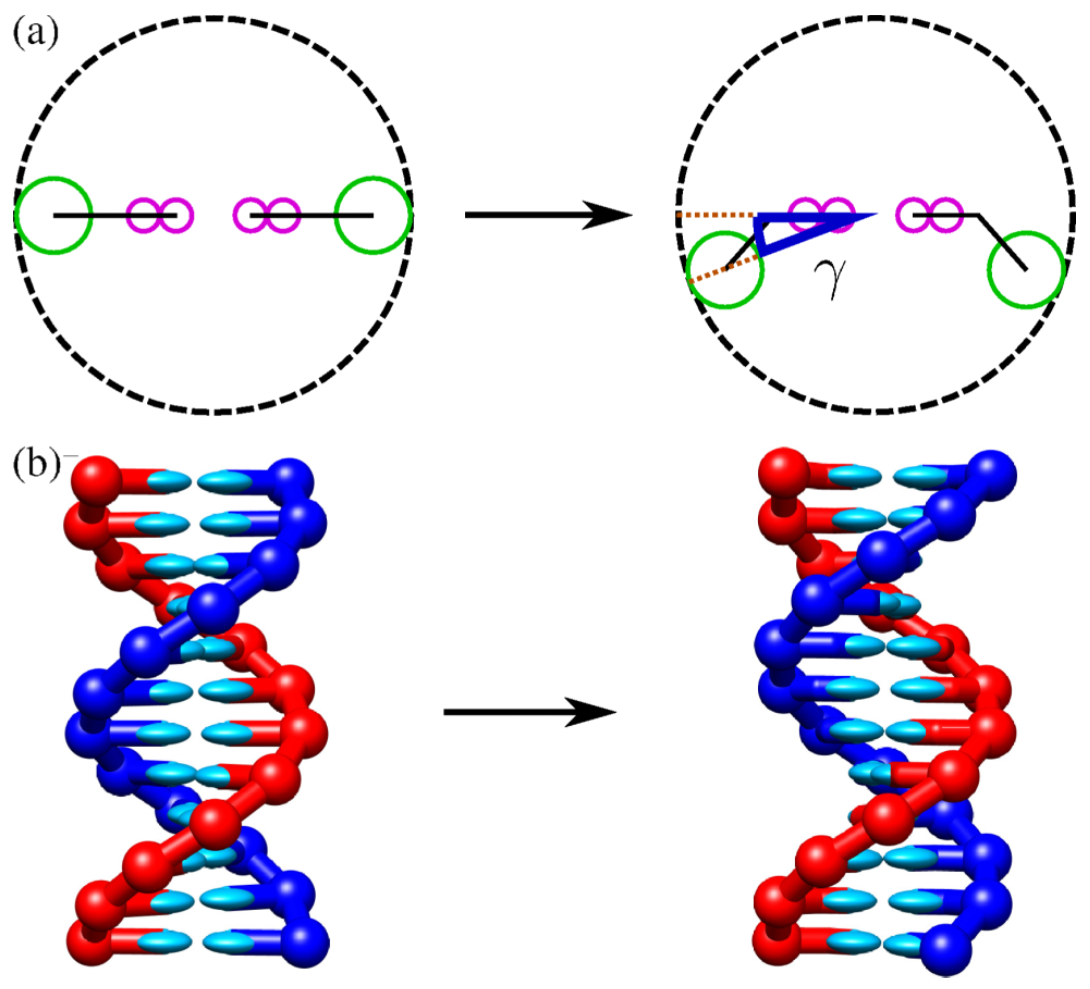}
\caption{\label{oxdna_oxdna2} (a): Schematic distinction between oxDNA (left) and oxDNA2 (right). In oxDNA all interaction sites are co-linear whereas in oxDNA2 the backbone interaction site and the 
stacking and hydrogen-bonding interaction sites are oriented at an angle. (b): The non-co-linear arrangement of the interaction sites leads to the formation of the major and minor groove, an
important structural feature of DNA. Reproduced from \cite{Snodin:2015}, with the permission of AIP Publishing.}
\end{center}
\end{figure}

The original model (oxDNA) has been further developed to include sequence-specific stacking and hydrogen bonding interaction strengths \cite{Sulc:2012} (oxDNA1.5) and implicit ions, which are modelled by means of a Debye-H\"uckel potential \cite{Snodin:2015} (oxDNA2).
A major improvement of the latest version is also the fact that it shows the correct structure with major and minor grooves (see Fig. \ref{oxdna_oxdna2} (b)). This is achieved through a modification of the relative position of the backbone and stacking/hydrogen bonding interaction sites, as schematically depicted in Fig. \ref{oxdna_oxdna2} (a).

%%%%%%%%%%%%%%%%%%%%%%%%%%%%%%%%%%%%%%%%%%%%%%%%%
% Implementation
%%%%%%%%%%%%%%%%%%%%%%%%%%%%%%%%%%%%%%%%%%%%%%%%%

\section{The LAMMPS Implementation of oxDNA}\label{implementation}

\subsection{Code Distribution, Force Fields and Compilation}

The software is open source and distributed under GNU General Public License (GPL).
It is available for download as LAMMPS USER-package from the central LAMMPS repository at 
Sandia National Laboratories, USA \cite{lammps}. This includes a detailed online documentation,
examples and utility scripts. We refer also to these materials for a general introduction into the
usage of LAMMPS.

To compile the code, load the LAMMPS standard packages \texttt{MOLECULE} and \texttt{ASPHERE} 
and the \texttt{USER-CGDNA} package by issuing\\ 
\texttt{make yes-molecule yes-asphere yes-user-cgdna}\\
in the main source code directory and compile as usual.  

All three versions oxDNA, oxDNA1.5 and oxDNA2 are implemented in the LAMMPS code and can be invoked through appropriate keywords in the input file.
This allows for instance to run without sequence-specific interactions and without implicit ions (oxDNA force field and keyword \texttt{seqav} $\equiv$ oxDNA), with sequence-specific interactions and without implicit ions (oxDNA force field and keyword \texttt{seqdep} $\equiv$ oxDNA1.5) or with implicit ions and with or without sequence-specific interactions (oxDNA2 force field and keywords \texttt{seqdep} or \texttt{seqav}, respectively).

The source code is also distributed via our main repository at 
CCPForge \cite{ccpforge} under the project name 
{\it Coarse-Grained DNA Simulation (cgdna)}.
Please send a request to join the project for full access that includes 
permission to browse the repository and commit changes. 

\subsection{Force and Torque Calculation}

Integrating the equations of motion of rigid bodies requires accurate information of their relative orientations. 
In simple situations this can be achieved through Euler angles, which 
describe the orientation of a rigid body and its local reference frame with respect to 
the laboratory system. Euler angles have the disadvantage that they are 
not unambiguously defined as a singularity arises when two rotation axes fall parallel. 
This situation, usually referred to as gimbal lock, arises easily in a system that 
contains a large number of rigid bodies. Unsurprisingly, it triggers numerical instabilities,
which is why rigid body problems are best formulated by means of quaternions \cite{AllenTildesley} 
instead of Euler angles.

Computationally it is most efficient to integrate the quaternion degrees of freedom directly via
a generalised 4-component quaternion torque (see \cite{tom-thesis} for a detailed derivation
of the oxDNA forces and generalised 4-torques using quaternion dynamics).
Unfortunately such an interface for generalised quaternion torques and momenta is not provided in LAMMPS.
It expects for its rigid body integrators 3-component torques and angular momenta as input quantities 
(besides the Newtonian force for the integration of the coordinate degrees of freedom). 
To be consistent and simplify interfacing with existing functionality, we decided to adhere to this convention. 
This, however, entails conversion of the unit quaternions into 
Cartesian unit vectors of a body frame before forces and torques can be calculated for the integration step,
thus leading to a computational overhead (see Appendix \ref{profiling}).

Once this choice has been made, the calculation of the forces and torques is most conveniently formulated
following Ref. \cite{Allen:2006}. If $\hat{\bm{a}}$ and $\hat{\bm{b}}$ are the principal axes of two rigid
bodies A and B and $r$ is the norm of the relative distance vector $\bm{r}=\bm{r}_A-\bm{r}_B$ from B to A,  
then the pair potential depends on a combination of these quantities,

\begin{equation}
U = U(r, \hat{\bm a},\hat{\bm b}) = U(r,\{\hat{\bm a}_m \cdot \hat{\bm r}\}, \{\hat{\bm b}_n \cdot \hat{\bm r}\}, \{\hat{\bm a}_m\cdot\hat{\bm b}_n\})
\end{equation}

\noindent where $\hat{\bm r}, \hat{\bm a}_m$ and $\hat{\bm b}_n$ are the normalised relative distance and orthonormal principal axes vectors.  
From this definition the force on A due to B are straightforwardly written as

\begin{eqnarray}
\bm{F}_A &=& -\bm{F}_B = -\frac{\partial U}{\partial \bm{r}}=\nonumber\\ 
&&\hspace*{-0.75cm} - \frac{\partial U}{\partial r}\hat{\bm r} - r^{-1}\sum_m\left[\frac{\partial U}{\partial (\hat{\bm{a}}_m\cdot\bm{r})}\hat{\bm{a}}_m^\perp + \frac{\partial U}{\partial (\hat{\bm{b}}_m\cdot\bm{r})}\hat{\bm{b}}_m^\perp\right].
\end{eqnarray}

\noindent Here $\hat{\bm{a}}_m^\perp = \hat{\bm{a}}_m - (\hat{\bm{a}}_m\cdot \hat{\bm r})\hat{\bm r}$ denotes the component
of $\hat{\bm{a}}_m$ which is perpendicular to $\hat{\bm r}$. The torques are slightly more involved:

\begin{eqnarray}
\bm{\tau}_A &=& \sum_m \frac{\partial U}{\partial (\hat{\bm{a}}_m\cdot\bm{r})}(\hat{\bm r}\times\hat{\bm{a}}_m)\nonumber\\
&& \hspace*{1.5cm}-\sum_{m n} \frac{\partial U}{\partial (\hat{\bm{a}}_m\cdot\hat{\bm{b}_n})}(\hat{\bm{a}}_m\times\hat{\bm{b}}_n)\\
\bm{\tau}_B &=& \sum_n \frac{\partial U}{\partial (\hat{\bm{b}}_n\cdot\bm{r})}(\hat{\bm r}\times\hat{\bm{b}}_n)\nonumber\\
&& \hspace*{1.5cm}+\sum_{m n} \frac{\partial U}{\partial (\hat{\bm{a}}_m\cdot\hat{\bm{b}_n})}(\hat{\bm{a}}_m\times\hat{\bm{b}}_n).
\end{eqnarray}
 
\noindent The fact that local angular momentum conservation requires 

\begin{equation}
\bm{\tau}_A + \bm{\tau}_B + \bm{r}\times\bm{f}=0
\end{equation}

\noindent can be conveniently utilised for debugging and verification purposes. The implementation was verified
against two independent implementations, namely Ouldridge's own code, which is based on
quaternion dynamics \cite{tom-thesis} as well as the standalone oxDNA code \cite{oxdna}, which makes also use of the
same scheme for the force and torque calculation. To this end two benchmarks were studied, 
a 5-base-pair duplex and a 8-base pair nicked duplex, which are both
provided as examples in the USER-CGDNA package.

\subsection{Input File}

In the following we discuss the structure of the input file and how the newly introduced oxDNA classes are invoked.

\noindent We work with Lennard-Jones reduced units, which are invoked in LAMMPS via

\smallskip
\noindent\texttt{units lj}
\smallskip

\noindent The system is three-dimensional.

\smallskip
\noindent\texttt{dimension 3}
\smallskip

\noindent In LAMMPS, an oxDNA nucleotide is represented as a bonded-ellipsoidal hybrid particle with the associated degrees of freedom of 
bonded particles in a bead-spring polymer (backbone connectivity) and aspherical particles 
with shape (moment of inertia), quaternion (orientation) and angular momentum.

\smallskip
\noindent\texttt{atom\_style hybrid bond ellipsoid}
\smallskip

\noindent Users are required to suppress the atom sorting algorithm as this can lead to problems in the bond topology of the DNA.

\smallskip
\noindent\texttt{atom\_modify sort 0 1.0}
\smallskip

\noindent It is important to set the skin size correctly, which controls the extent of the neighbour lists. Too large a skin size and neighbour lists
become unnecessarily long, leading to superfluous communication. Too short and partners in the pair interactions will be
lost. 

\smallskip
\noindent\texttt{neighbor 1.0 bin}
\smallskip

\noindent A good way to fine-tune this parameter is to run an NVE simulation with constant energy before applying Langevin integrators.
We recommend \texttt{neighbor 2.0 bin} as a safe starting point. Likewise, frequent update of the neighbour lists can lead to an undue performance degradation. This parameter should be tuned 
as well so that no dangerous builds (as reported in the standard output of LAMMPS) occur.

\smallskip
\noindent\texttt{neigh\_modify every 1 delay 0 check yes}
\smallskip

\noindent The initial configuration and topology is created by means of an external setup tool (see Sec. \ref{data_setup}) and read in.

\smallskip
\noindent\texttt{read\_data data\_file\_name}
\smallskip

\noindent All masses are set to $3.1575$ in LJ units.

\smallskip
\noindent\texttt{set atom * mass 3.1575}
\smallskip

\noindent Note that the moment of inertia is determined through the shape parameter in the data file (see below Sec. \ref{data_setup}). 
There are four types of nucleotides (A=1, C=2, G=3, T=4), which are grouped together into a group named \texttt{all} for the integration. 

\smallskip
\noindent\texttt{group all type 1 4}
\smallskip

\noindent The new oxDNA classes with its parameters are invoked as follows:

\smallskip
\noindent \texttt{bond\_style oxdna2/fene\\
\smallskip
bond\_coeff * 2.0 0.25 0.7564\\
pair\_style hybrid/overlay oxdna2/excv \&\\
\hspace*{0.5cm} oxdna2/stk oxdna2/hbond oxdna2/xstk \&\\
\hspace*{0.5cm} oxdna2/coaxstk oxdna2/dh\\
pair\_coeff * * oxdna2/excv   2.0 0.7 0.675 2.0 \&\\
\hspace*{0.5cm} 0.515 0.5 2.0 0.33 0.32\\
pair\_coeff * * oxdna2/stk    seqdep 0.1 6.0 0.4 \&\\
\hspace*{0.5cm} 0.9 0.32 0.6 1.3 0 0.8 0.9 0 0.95 0.9 0 \&\\
\hspace*{0.5cm} 0.95 2.0 0.65 2.0 0.65\\   
pair\_coeff * * oxdna2/hbond  seqdep 0.0 8.0 \&\\
\hspace*{0.5cm} 0.4 0.75 0.34 0.7 1.5 0 0.7 1.5 0 0.7 1.5 \&\\
\hspace*{0.5cm} 0 0.7 0.46 3.141592653589793 0.7 4.0 \&\\
\hspace*{0.5cm} 1.5707963267948966 0.45 4.0 \&\\
\hspace*{0.5cm} 1.5707963267948966 0.45\\
pair\_coeff 1 4 oxdna2/hbond  seqdep 1.0678 8.0 \&\\
\hspace*{0.5cm} 0.4 0.75 0.34 0.7 1.5 0 0.7 1.5 0 0.7 1.5 \&\\
\hspace*{0.5cm} 0 0.7 0.46 3.141592653589793 0.7 4.0 \&\\
\hspace*{0.5cm} 1.5707963267948966 0.45 4.0 \&\\
\hspace*{0.5cm} 1.5707963267948966 0.45\\
pair\_coeff 2 3 oxdna2/hbond  seqdep 1.0678 8.0 \&\\
\hspace*{0.5cm} 0.4 0.75 0.34 0.7 1.5 0 0.7 1.5 0 0.7 1.5 \&\\
\hspace*{0.5cm} 0 0.7 0.46 3.141592653589793 0.7 4.0 \&\\
\hspace*{0.5cm} 1.5707963267948966 0.45 4.0 \&\\
\hspace*{0.5cm} 1.5707963267948966 0.45\\
pair\_coeff * * oxdna2/xstk  47.5 0.575 0.675 \&\\
\hspace*{0.5cm} 0.495 0.655 2.25 0.791592653589793 0.58 \&\\
\hspace*{0.5cm} 1.7 1.0 0.68 1.7 1.0 0.68 1.5 0 0.65 1.7 \&\\
\hspace*{0.5cm} 0.875 0.68 1.7 0.875 0.68\\ 
pair\_coeff * * oxdna2/coaxstk 58.5 0.4 0.6  \&\\
\hspace*{0.5cm} 0.22 0.58 2.0 2.891592653589793 0.65 1.3 \&\\
\hspace*{0.5cm} 0 0.8 0.9 0 0.95 0.9 0 0.95 40.0 \&\\
\hspace*{0.5cm} 3.116592653589793\\
pair\_coeff * * oxdna2/dh      0.1 1.0 0.815\\
}

\noindent Please note that according to the LAMMPS parsing rules the ampersands (\&) represent line breaks.\\
Visit the LAMMPS online documentation and manual for more information and for information on oxDNA2.

\subsection{Data File and Setup Tool}\label{data_setup}

The data file contains all relevant structural parameters for the simulation, i.e. details about 
the number of atoms, the topology of the molecules, the size of the simulation box, initial velocities, etc. 
The LAMMPS implementation of oxDNA follows the standard form as discussed in the LAMMPS user manual. 
We outline the relevant parts below.\\

\noindent At the beginning of the data file the total number of particles and bonds has to be given. As we are using
hybrid particles, we need to set the same number of ellipsoids. For a standard DNA duplex consisting
of 8 complementary base pairs we need 16 atoms, 16 ellipsoids and 14 bonds, 7 on each of the two single strands.
If the strands are nicked, which we do not assume here, the number of bonds would be reduced. 

\smallskip
\noindent\texttt{16 atoms\\
16 ellipsoids\\
14 bonds
}
\smallskip

\noindent We use four atom types to represent the four different nucleotides in DNA (A=1, C=2, G=3, T=4). 
We use only one bond type.

\smallskip
\noindent\texttt{4 atom types\\
1 bond types
}
\smallskip

\noindent The dimensions of the simulation box are defined as follows:

\smallskip
\noindent\texttt{-20.0 20.0 xlo xhi\\
-20.0 20.0 ylo yhi\\
-20.0 20.0 zlo zhi
}
\smallskip

\noindent Although already stated in the input file, we need to provide again the masses of the nucleotides.

\smallskip
\noindent\texttt{Masses\\
1 3.1575\\
2 3.1575\\
3 3.1575\\
4 3.1575
}
\smallskip

\noindent The nucleotides are defined after the keyword \texttt{Atoms}.
Each row contains the atom-ID (1,2,3 in the example below), the atom type (1,1,4), the position (x,y,z), 
the molecule ID (all 1 in this case), an ellipsoidal flag (1) and a density (1). 

\smallskip
\noindent\texttt{Atoms\\
1  1  0.00000 \ 0.00000  0.00000  1 1 1\\
2  1  0.13274  -0.42913  0.37506  1 1 1\\
3  4  0.48461  -0.70835  0.75012  1 1 1\\
$\vdots$
}
\smallskip

\noindent Next we set the initial velocities to the desired value, here all equal to 0.
The first column contains the atom-ID (1,2,3), the following three columns the translational,
and the last three columns the angular velocity.

\smallskip
\noindent\texttt{Velocities\\
1  0.0  0.0  0.0  0.0  0.0  0.0\\ 
2  0.0  0.0  0.0  0.0  0.0  0.0\\ 
3  0.0  0.0  0.0  0.0  0.0  0.0\\
$\vdots$
}
\smallskip

\noindent Note that this is our special choice in the setup tool. The velocities can be generally initialised to any value.
Large values will lead to the FENE springs becoming overstretched and may provoke an early abortion of the run. 

\noindent The ellipsoids are defined with atom-ID, shape (1.17398 to produce the correct moment of inertia) 
and initial quaternion (last four columns). 

\smallskip
\noindent\texttt{Ellipsoids\\
1   1.17398 1.17398 1.17398  1.00000  0.  0.  0.\\
2   1.17398 1.17398 1.17398  0.95534  0.  0.  0.29552\\
3   1.17398 1.17398 1.17398  0.82534  0.  0.  0.56464\\
$\vdots$
}
\smallskip

\noindent Finally, we specify the bond topology. The first column contains the bond-ID (1,2,3),
the second one the bond type (1) and the third and fourth the IDs of the two bond partners.

\smallskip
\noindent\texttt{Bonds\\
1       1       1       2\\
2       1       2       3\\
3       1       3       4\\
$\vdots$
}
\smallskip

\noindent To simplify the setup procedure we provide a simple python tool with the 
example and utility files of the USER-CGDNA package. The script allows the user to create 
single- and double-stranded DNA from an input file that specifies the sequence and
requires an installation of \texttt{numpy}.

\noindent The syntax is very straightforward, but the system size has to be 
specified in the following way:

\smallskip
\noindent\texttt{\$> python generate.py <box\_offset> $\backslash$\\ 
\hspace*{0.75cm} <cubic\_box\_length> <sequence\_file\_name>
} 
\smallskip

\noindent The output is written directly into a data file in LAMMPS format. This has to 
be given in the LAMMPS input file. \texttt{<sequence\_file\_name>} is an ASCII input file 
that contains keywords and the sequence of one ssDNA strand. Two options are available.
For a single, helical strand consisting of ssDNA, the sequence file contains a single line: 

\smallskip
\noindent\texttt{ACGTA}
\smallskip

\noindent If the sequence is prepended by the keyword \texttt{DOUBLE}, then 
a single, helical DNA duplex is created. The bases on the second strand are complementary 
to those on the first strand, which is given in the sequence input file: 

\smallskip
\noindent\texttt{DOUBLE ACGTA}
\smallskip

\noindent Consecutive strands are positioned and oriented randomly without creating any overlap in case of more than one ssDNA or dsDNA strand. 
Note that the procedure works only below a critical density as this simple script does not feature cell lists.
\com{Besides these setup tools, the USER-CGDNA package contains as well example input, data and standard output files 
of short benchmark runs of dsDNA duplexes.}

\subsection{Output and Visualisation}

LAMMPS offers a multitude of possible output formats, including parallel HDF5 and NetCDF formats,
VTK format or very basic standard trajectory data. We will summarise here how output of basic 
observables of the oxDNA model can be invoked in the input file. 

The xyz style writes XYZ files, which is a simple text-based coordinate format that many codes can read,
which has one line per atom with the atom type and the x-, y-, and z-coordinate of that atom. This
style is invoked via

\smallskip
\noindent\texttt{dump 1 all xyz Nint trajectory.xyz}
\smallskip

\noindent where \texttt{Nint} is the output frequency in timesteps. 
Additional output of e.g. velocity, force and torque on a per-atom basis makes
some customisation necessary,

\smallskip
\noindent\texttt{dump 2 all custom Nint filename.dat id x y z \&\\
\hspace*{0.5cm} vx vy vz fx fy fz tqx tqy tqz}
\smallskip

where \texttt{id} is the unique atom-ID. The output of quaternions requires 
a so-called \texttt{compute} style. The result of the \texttt{compute} style can then be 
retrieved in the following way:

\smallskip
\noindent\texttt{compute quat all property/atom quatw quati \&\\
\hspace*{0.5cm} quatj quatk\\
dump 3 all custom Nint filename.dat id \&\\
\hspace*{0.5cm} c\_quat[1] c\_quat[2] c\_quat[3] c\_quat[4]
}
\smallskip

Another observable that may be of interest is the energy,
or more specifically broken down into rotational, kinetic and 
potential energy. This is also done through a \texttt{compute}
style.

\smallskip
\noindent\texttt{compute erot all erotate/asphere\\
compute ekin all ke\\
compute epot all pe\\
variable erot equal c\_erot\\
variable ekin equal c\_ekin\\
variable epot equal c\_epot\\
variable etot equal c\_erot+c\_ekin+c\_epot
}
\smallskip

Note that the somewhat simpler \texttt{thermo\_style} command for output 
discards the kinetic energy of rotation when the kinetic energy is requested.

LAMMPS does not contain a direct visualisation toolkit. There are, however,
a multitude of ways how snapshots can be visualised.  ParaView \cite{paraview} for instance,
is an open source, multi-platform data analysis and visualisation application. 
The images in this work have been generated with the molecular visualisation program 
VMD (Visual Molecular Dynamics) \cite{VMD}. More information about possible 
visualisation pipelines can be found in the LAMMPS online
manual \cite{lammps}.

%%%%%%%%%%%%%%%%%%%%%%%%%%%%%%%%%%%%%%%%%%%%%%%%%
% Rigid-body integrators
%%%%%%%%%%%%%%%%%%%%%%%%%%%%%%%%%%%%%%%%%%%%%%%%%

\section{Langevin-Type Rigid-Body Integrators}\label{integrator}

Together with the USER-CGDNA package comes also an implementation of novel Langevin-type rigid-body integrators
 that were developed by Davidchack, Ouldridge and Tretyakov \cite{Davidchack:2015}.
The motivation for this was that previously only a limited choice of suitable Langevin integrators for rigid bodies
was available in LAMMPS.
Without noise all integrators A, B and C in the above reference are identical and basically equivalent to the integrator presented 
by Miller et al. \cite{Miller:2002}.
Nevertheless, we refer to this case as the ``DOT integrator'' (the other implementation of the Miller integrator 
is only available when using the \texttt{fix rigid} command in LAMMPS). 
The DOT integrator is an alternative to the  
standard LAMMPS NVE integrator for aspherical particles, and can be invoked by replacing
the standard choice

\smallskip
\noindent\texttt{fix 1 all nve/asphere}
\smallskip

\noindent with

\smallskip
\noindent\texttt{fix 1 all nve/dot}
\smallskip

\noindent in the input file. This energy-conserving integrator is useful for an analysis of 
the accuracy of this family of integrators or the integrity of the pair interactions at a given timestep size 
$\Delta t$.

The C integrator in Ref. \cite{Davidchack:2015}, to which we refer as ``DOT-C integrator'',
 is invoked by replacing the standard NVE integrator 
for aspherical particles and the fix for Langevin dynamics 

\smallskip
\noindent\texttt{fix 1 all nve/asphere}\\
\texttt{fix 2 all langevin 0.1 0.1 0.03 457145 angmom 10}
\smallskip

\noindent with one single fix

\smallskip
\noindent\texttt{fix 1 all nve/dotc/langevin 0.1 0.1 0.03 \&\\
\hspace*{0.75cm} 457145 angmom 10}
\smallskip

\begin{figure}[htpb]
\begin{center}
\includegraphics[width=0.5\textwidth]{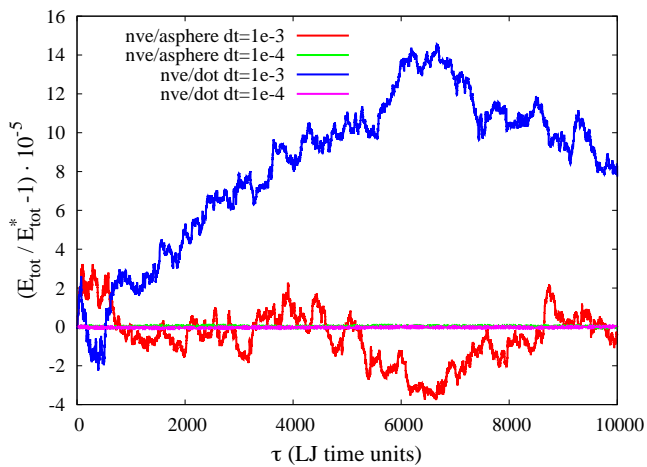}
\caption{\label{integrator_comparison} Relative normalised accuracy $(E_{tot}-E^*_{tot})/E^*_{tot}$ of the standard LAMMPS NVE integrator 
for aspherical particles and the NVE DOT integrator from Ref. \cite{Davidchack:2015}. $E^*_{tot}$ is the total free energy 
at the beginning of the simulation runs.}
\end{center}
\end{figure}

To measure the accuracy of the new integrators, we run a test case consisting of a short, 
nicked duplex with 8 base pairs (16 nucleotides).
Fig. \ref{integrator_comparison} shows the accuracy measured through the normalised difference between the total energy $E_{tot}$ for
this particular benchmark and the total energy at the beginning of the run $E_{tot}^*$. 
We compared the standard \texttt{fix nve/asphere} integrator, 
which is based on a Richardson iteration in the update of the quaternion degrees of freedom, to the new DOT integrator, 
which uses a rotation sequence to update the quaternions. Shown are results for two different timestep sizes
$\Delta t=10^{-3}$ and $\Delta t=10^{-4}$. Both simulations were run for the same physical simulation time to allow
 direct comparison of the deviations of a dynamical run. As this is done in the NVE ensemble and without noise, 
the energy should be exactly conserved. This corresponds to a straight, horizontal line at 0.

It is obvious that above a certain timestep size the accuracy of the new DOT integrator is slightly 
inferior compared to the standard integrator. Up to a certain point the DOT integrator actually seems to deviate 
further from the correct result, whereas the standard integrator fluctuates more around the correct value. 
This, however, is more or less a transient effect as longer runs show there is no permanent drift away from the correct
result. 
\begin{table*}[htpb]
\begin{center}
\begin{tabular}{ | c | c | c | c | c | c | c |}
\hline
& $\Delta t$ & $E_{kin}$ & $E_{rot}$ & $E_{pot}$ & $E_{tot}$ & standard error of $E_{tot}$ fit\\
\hline
\hline
fix nve/asphere \& & & & & & &\\
fix langevin & & & & & &\\
\hline
& $10^{-4}$ &2.3999 &2.4001  &-21.4512 &-16.6513 & $\pm$ 0.00377  (0.0227\%)\\
\hline
& $10^{-3}$ & 2.4015  &2.4021  &-21.5564 &-16.7582 & $\pm$ 0.00349 (0.0208\%)\\
\hline
& $5\cdot10^{-3}$ &2.4012  &2.3999  &-21.6352  &-16.8315  & $\pm$ 0.00322 (0.0191\%)  \\
\hline
\hline
nve/dotc/langevin & & & & &\\
\hline
& $10^{-4}$ &2.3989 &2.3997 &-21.5278 &-16.7292 & $\pm$ 0.00362 (0.0216\%)\\
\hline
& $10^{-3}$ &2.3998 &2.4008 &-21.6631 &-16.8624 & $\pm$ 0.00335 (0.0199\%)\\
\hline
& $10^{-2}$ & 2.3959 & 2.3941  &-21.6151 &-16.8251 & $\pm$ 0.00318 (0.0189\%) \\
\hline
& $2\cdot10^{-2}$ &2.3895  & 2.3752 &-21.6266   &-16.8619  & $\pm$ 0.00313 (0.0185\%)\\
\hline
\end{tabular}
\end{center}
\caption{\label{table1} Average kinetic, rotational, potential and total energy for the standard LAMMPS integrator \texttt{fix nve/asphere} \&
\texttt{fix langevin} and the DOT-C integrator \texttt{nve/dotc/langevin} for different timestep sizes.}
\end{table*}   
For Langevin dynamics, it is not possible to evaluate the accuracy and stability in the same way.
We opted instead for an estimate based on the average kinetic, rotational, potential and
total energy of the benchmark. Again, we performed runs of $\tau=10000$ Lennard-Jones time units length, this time thermalised, 
and averaged the results over the time interval. The number of MD-timesteps and the output frequency for each timestep size 
were adapted so that the total physical simulation time and the statistical basis of the error calculations were consistent. 
The temperature in reduced LJ-units was set to $T=0.1$, whereas the translational and rotational friction or damping coefficients 
were set to $\gamma=1/0.03$ and $\Gamma=1/0.3$, respectively. The results are summarised in Tab. \ref{table1}. 
These values were used during the verification of the LAMMPS implementation because they produced relatively smooth trajectories that could be easily followed. For actual production runs it may be more appropriate to use different values to allow a better and more efficient sampling of the configuration space.

Based on three translational and three rotational degrees of freedom per nucleotide and 8 base pairs we expect kinetic and rotational energies $E_{kin}=E_{rot}=2.4$ 
for a temperature settings $T=0.1$. This is very well achieved for all timestep sizes and both integrators, the standard LAMMPS integrator \texttt{fix nve/asphere \& fix langevin} and the DOT-C integrator \texttt{fix nve/dotc/langevin}. However, 
there appears to be a slight decrease in the 
DOT-C integrator for very large step sizes ($\Delta t=2\cdot10^{-2}$).
The deviation of the total energy between all timestep sizes, admittedly an {\it ad hoc} criterion to quantify the stability of the integrators,
but one that is rather hard for the integrators to get exactly right, is in the sub-percent range.
It is actually slightly better for the DOT-C integrator than for the standard LAMMPS integrator. 
The statistical errors, reported in Tab.\ref{table1}, are the standard deviations of a linear least square fit and 
show that the deviations are well above the uncertainty of the fits.

Remarkably, for the DOT-C integrator the limit for a stable integration is $\Delta t=2\cdot10^{-2}$, which represents a very large timestep size. 
This is about 4 times larger than the maximum timestep size for which the standard LAMMPS Langevin integrator produces sound results.
Because of the more complex rotations in quaternion space and various additional transformations that the DOT-C integrator requires
there is a small overhead of about $15\%$ compared to the standard LAMMPS integrator. Nevertheless, this small overhead of the DOT-C
integrator is very well compensated by the computational efficiency and possibility to increase the timestep 
size by $400\%$ (from a maximum of $\Delta t=5\cdot10^{-3}$ for the standard LAMMPS integrator to $\Delta t=2\cdot10^{-2}$ for the DOT-C integrator).

%%%%%%%%%%%%%%%%%%%%%%%%%%%%%%%%%%%%%%%%%%%%%%%%%
% Performance
%%%%%%%%%%%%%%%%%%%%%%%%%%%%%%%%%%%%%%%%%%%%%%%%%

\section{Performance Analysis}\label{performance}

We devised a few simple benchmarks to study the parallel performance of the LAMMPS implementation. The size of each benchmark is well beyond the 
current capabilities of the standalone version, so each demonstrates as well a minimal performance requirement. 
\begin{figure}[htpb]
\begin{center}
\includegraphics[width=0.235\textwidth]{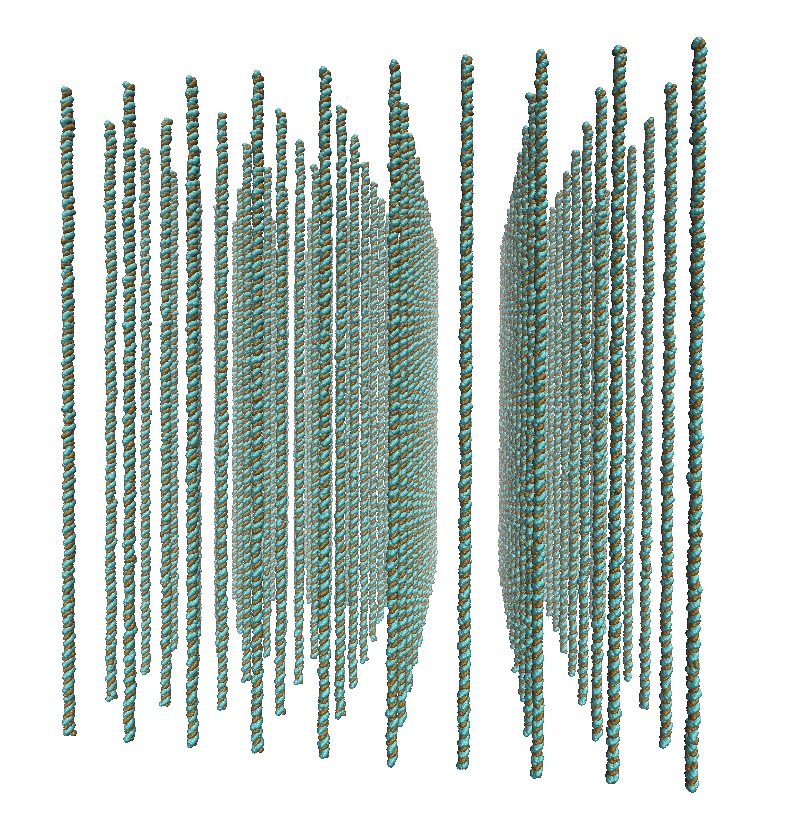}
\includegraphics[width=0.235\textwidth]{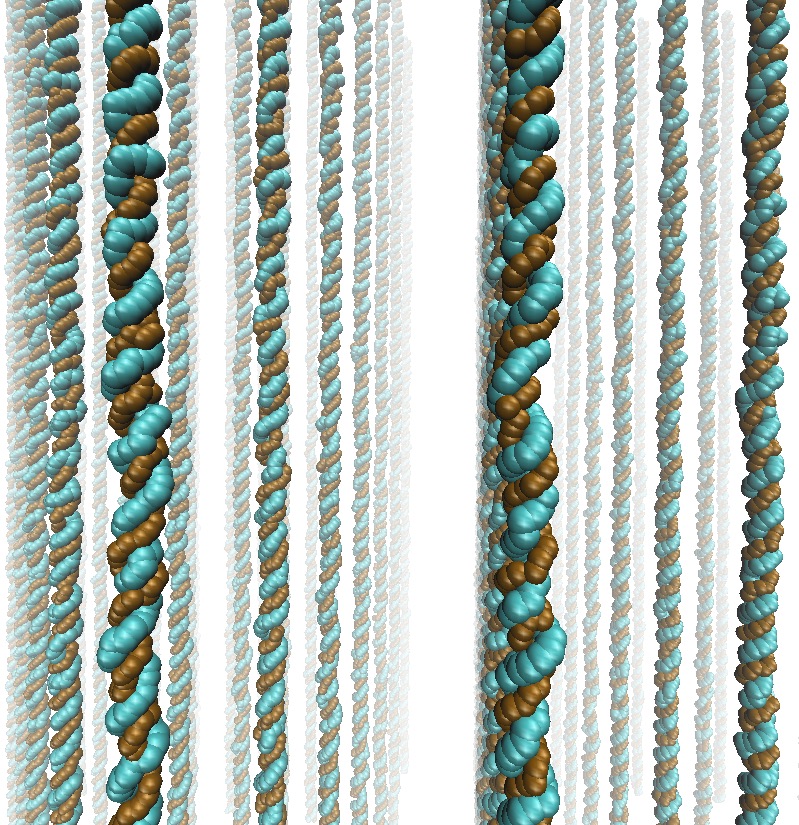}
\caption{\label{bench-60kbps} The low-density benchmark consisting of a $10\times 10$ array of DNA duplexes with A-T base pairs and a length of 600 base pairs each, in total 60 kbp. The high-density benchmark (not shown) consisted of a similar $40\times 40$ array of duplexes with 960 kbp in total. The pictures show the final configuration the end of a performance run and were produced with VMD. \com{The centre of mass of each nucleotide is represented through a sphere.}}
\end{center}
\end{figure}
The benchmarks consisted of arrays of double-stranded, regularly arranged DNA duplexes, each with a length of 600 base pairs.
The low-density (LD) benchmark was formed by a $10 \times 10$ array of duplexes, giving a total of 60 kbp, and is shown in Fig. \ref{bench-60kbps}. 
The high-density (HD) benchmark was formed by a $40\times 40$ array of duplexes with a density 16 times larger than the LD case and a total number
of 960 kbp.
\begin{figure}[htpb]
\begin{center}
\includegraphics[width=0.5\textwidth]{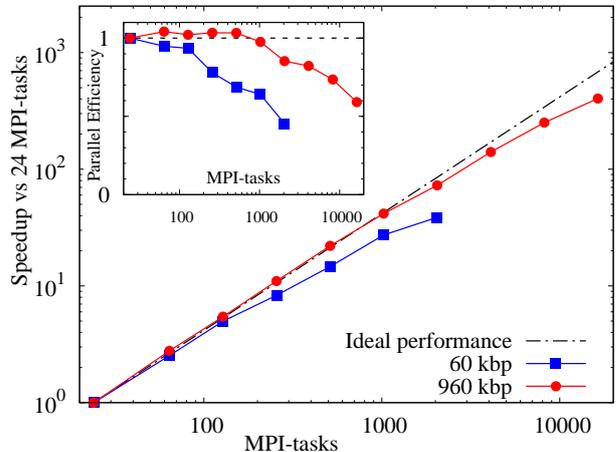}
\caption{\label{scaling} Strong scaling behaviour: Speedup of the low and high density benchmarks of 60 kbp and 960 kbp, respectively, compared
to the single node performance with 24 MPI-tasks. The inset shows the parallel efficiency relative to the single node case with 24 MPI-tasks.}
\end{center}
\end{figure}
Whilst a regular array of double-stranded DNA strands appears perhaps somewhat artificial, it creates a reasonably load-balanced 
situation and facilitates the performance analysis. The obtained densities of DNA, are however very well comparable to 
those of DNA gels \cite{Rovigatti:2014} and high density states of DNA which form liquid-crystalline phases \cite{DeMichele:2012}.

Strong scaling tests were performed on ARCHER on up to 86 nodes (LD) and 683 nodes (HD), respectively.
The benchmark cases were run for 30,000 (LD) and 10,000 (HD) MD-timesteps with a timestep size of $\Delta t=5\times10^{-3}$.
We used the standard LAMMPS integrators for Langevin dynamics, although the scaling behaviour was found to be virtually
identical when using the above described rigid body integrator DOT-C.
The primary reason for this was that the wallclock time for runs with the standard integrator was still a few
percent shorter, although the improved efficiency of the DOT-C integrator would mean these runs were shorter in physical time. 
The temperature in reduced LJ-units was $T=0.1$, whereas the translational and rotational friction coefficients were set to 
$\gamma=1/0.03$ and $\Gamma=1/0.3$, respectively.

Fig. \ref{scaling} shows the parallel speedup for both benchmarks relative to the single node performance with 24 MPI-tasks. The code performs well
for the LD benchmark up to about 128 MPI-tasks with a parallel efficiency around $95\%$ (see the inset). Beyond several hundred MPI-tasks a gradual performance 
degradation is observed. At 2048 MPI-tasks the parallel efficiency has decreased to about $45\%$ and the total speedup is roughly 930-fold compared to the single core performance (39-fold compared to the single node performance).\\
A look at the ratio of the number of local atoms, i.e. those that are inside a process boundary, to the number of ghost atoms, i.e. those which need to be communicated via neighbour lists, 
proves that the observed performance degradation is due to the comparably small size of the problem. 
At the largest core counts there are on average only about 60 local atoms present on each process, whereas the 
number of ghost atoms is with about 225 atoms almost four times larger. 
LAMMPS is known to require at least a few hundred local atoms or more for a good parallel performance \cite{LAMMPSdoc}. 
The speedup is still relatively good because the fraction of time that the algorithm spends in the force calculation is still comparably large.
For the HD benchmark, 16 times larger than the LD case, the performance degradation is more or less mirrored at core counts that are 
about 16 times larger. For the HD benchmark the total speedup at 16384 MPI-tasks is 9680-fold with respect to the single core performance 
(400-fold compared to the single node performance) and the parallel efficiency is still at around  $60\%$.

These two examples are of course slightly idealised in the sense that both benchmarks fulfil easily the requirement of good load-balancing,
which is necessary to obtain a good scaling performance. LAMMPS, however, features sophisticated load-balancing algorithms which permit good scaling behaviour also for very inhomogeneous systems.
We are planning to extend the existing implementation to benefit further from recent developments pertaining to threaded parallelisation on 
shared memory architectures such as many-core chips and general purpose graphical processing units (GPGPUs).

\com{One of the major advantages of the new LAMMPS implementation is that it can be directly compared with other 
coarse-grained models that are also based on the LAMMPS code. 
To this end, we compared the single core performance of oxDNA2 with that of 3SPN.2 \cite{Hinckley:2013}.
The benchmark consisted of two complementary dsDNA duplexes of 8 bps with implicit ions.
In order to compare both models we set the translational friction coefficient $\gamma$ to about (300\, fs)$^{-1}$.
We opted for the maximum timestep size that provided a stable integration, which was 
$\Delta t = 35$\, fs (3SPN.2) and $\Delta t = 48$\, fs (oxDNA2 + DOT-C integrator), respectively.}

\com{On a single Intel Core i7 2.8 GHz processor using the latest version
of LAMMPS (16 March 2018) 3SPN.2 delivered a performance of about 60 $\mu$s per day.
oxDNA2 was able to surpass this by about a factor 1.6 with a performance of roughly 100 $\mu$s per day.
Note that comparing the wall times is only an approximate way to compare the performance 
as there is no guarantee that similar processes take a similar simulation time in the two models.}

\com{Apart from the enhanced stability of the rigid body integrator, this difference in performance will be 
caused by the different number of degrees of freedom that both models require: oxDNA/oxDNA2 uses only 
13 degrees of freedom per nucleotide 
(3 coordinate positions, 3 translational momenta, 3 angular momenta and 4 quaternion degrees of freedom), 
whereas 3SPN.2 uses 18 degrees of freedom per nucleotide 
(3 particles with each 3 coordinate positions and translational momenta).} 

\com{Unfortunately, we could not measure the parallel performance of
3SPN.2. But this conceptual difference between the two models is very likely to entail further detrimental 
effects when running in parallel. With the larger number of degrees of freedom per nucleotide in 3SPN.2, 
communication overheads are likely to build up more quick\-ly and neighbour lists are longer and probably 
have to be rebuilt more frequently. 
On the other hand, the current LAMMPS implementation of oxDNA offers further 
potential for optimisation as it spends a good part its time computing the inverse cosine 
(around 12\%, see Appendix \ref{profiling}). This could be alleviated for instance through the 
introduction of appropriate lookup tables for trigonometric functions.}

%%%%%%%%%%%%%%%%%%%%%%%%%%%%%%%%%%%%%%%%%%%%%%%%%
% Applications
%%%%%%%%%%%%%%%%%%%%%%%%%%%%%%%%%%%%%%%%%%%%%%%%%

\section{Applications}\label{applications}

The structural properties of DNA such as the per\-sis\-tence length, radius of gyration and torsional rigidity play an important role in its function. Characterising these properties and their dependence on different conditions is therefore fundamental for highly complex processes such as DNA packaging, replication and denaturation.
Experimentally, however, making these measurements is not an easy task as it requires subtle manipulation of single molecules and direct measurement of their response to applied forces or displacements, which can then be related to the elasticity of DNA.
By using coarse-grained computational models like oxDNA, we can study these systems in more detail. These simulations can in turn provide insights into experimental data or the performance of other theoretical approaches.

%The radius of gyration is a particularly useful descriptor of the structure and compactness of macromolecules. %Since this physical quantity depends on the DNA topology \cite{DNAdifftopology,lambdascattering}, it is important to distinguish between three fundamental forms in which it may exist: .
%For ssDNA the radius of gyration $R_g$ can be defined as  

The radius of gyration is a particularly useful descriptor of the structure and compactness of macromolecules. For ssDNA the radius of gyration $R_g$ can be defined as

\begin{equation}
R^2_g = \frac{1}{N}\sum_{i=1}^N (\bm{r}_i-\bar{\bm{r}})^2
\end{equation}

\noindent where $N$ is the number of nucleotides, $\bm{r}_i$ is the position of the $i-th$ nucleotide and $\bar{\bm{r}}=\frac{1}{N}\sum_{i=1}^N \bm{r}_i$ is the mean position of the ssDNA strand. For dsDNA this definition would be modified to use the centre of mass coordinate of a base pair (bp) and $N$ would be replaced with the number of base pairs. 

In this section we present results obtained with the oxDNA2 model for two different systems: a sequence of ssDNA from a $\lambda$-bacteriophage that has a multitude of applications in microbial and molecular genetics and serves e.g. as cloning vector, as well as complete ssDNA sequence of the pUC19 plasmid, another model organism and cloning vector, which conveys antibiotic resistance. 

We performed Langevin dynamics simulations of the two above mentioned ssDNA sequences at a constant salt concentration of $0.2$M NaCl. For simplicity we used linear DNA molecules, so their ends are freely to rotate. After a sudden quench in temperature, the system evolved from a random initial configuration towards a new steady state. The criterion for reaching this steady state was a constant radius of gyration $R_g$ and number of base-pairs $N_{\rm c}$ formed along the chain. Equilibrium values for these observables were obtained by averaging five different configurations over the last $3\times 10^5$ $\tau_{LJ}$ timesteps.

\begin{figure}[htpb]
\begin{center}
\includegraphics[width=0.5\textwidth]{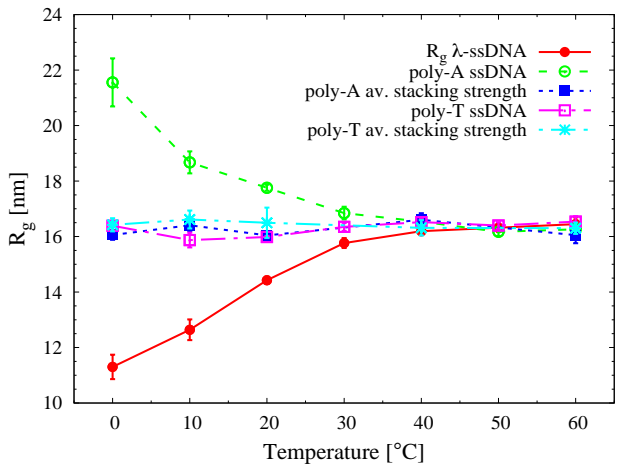}
\caption{\label{Rgvst500bp} Response of the radius of gyration $R_g$ of the ssDNA $\lambda$-DNA sequence to temperature changes. Points show $R_g$ computed from averages over various configurations of a 500 base pair (bp) long ssDNA chain using the oxDNA2 model with sequence-specific stacking strength (red full circles), poly-A (green open circles) and poly-T (magenta open squares). These results are compared to those for poly-A and poly-T chains with sequence-averaged stacking strength, respectively (blue full squares and cyan crosses).}
\end{center}
\end{figure}

\begin{figure}[htpb]
\begin{center}
\includegraphics[width=0.5\textwidth]{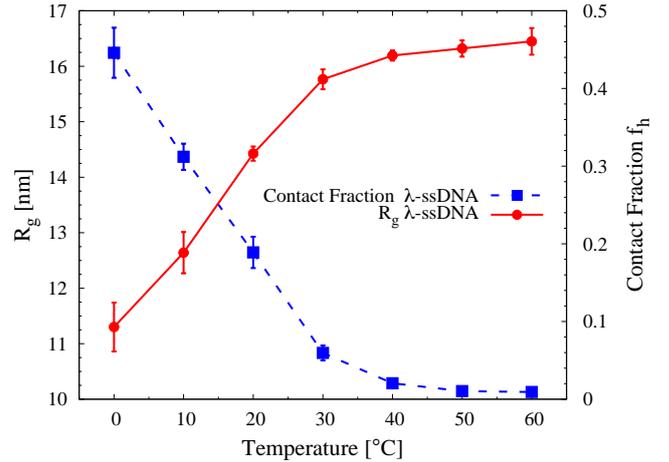}
\caption{\label{rg_contacts_ldna} Dependence of the radius of gyration $R_g$ and the fraction of direct intra-chain nucleotide contacts in the $\lambda$-ssDNA sequence on the temperature.
}
\end{center}
\end{figure}

In Fig.\ref{Rgvst500bp} the initial 500 nucleotide long sequence of ssDNA $\lambda$-DNA is compared with different linear DNA mo\-lecules of the same length, namely poly-A and poly-T strands. The radius of gyration as a function of temperature is shown. For $\lambda$-DNA we observe that $R_g$ increases with temperature until a plateau is reached at around 50$\degree$C. 
While the $\lambda$-DNA sequence allows hybridisation along the ssDNA (see Fig.\ref{rg_contacts_ldna}), the same is not true for poly-A or ploy-T sequences. This can explain the differences in $R_g$ between the two that we observe at low temperatures. 
In contrast, poly-A shows the opposite tendency, with the largest $R_g$ at the lowest temperature setting of 0$\degree$C. The reason for this different behaviour is the roughly 16\% larger stacking strength between consecutive A nucleotides as compared to T nucleotides, an explanation that is corroborated through a sequence-averaged stacking strength (see poly-A-avstk and poly-T-avstk). Finally, for higher temperatures self-hybridisation becomes less important and the radius of gyration approaches the same plateau value for all sequences.

\begin{figure}[htpb]
\begin{center}
\includegraphics[width=0.5\textwidth]{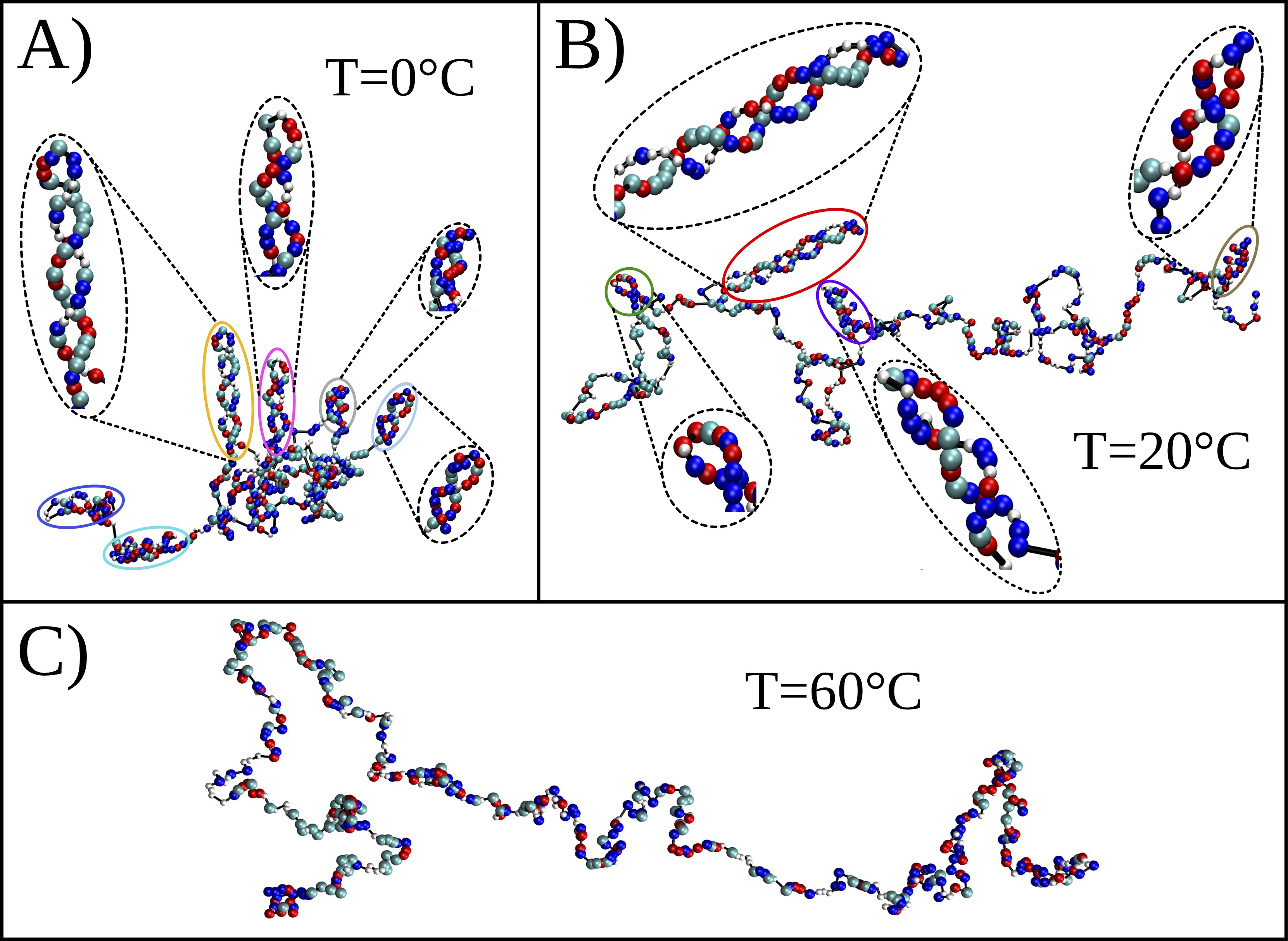}
\caption{\label{snapshots_ldna} Simulation snapshots of the $\lambda$-ssDNA sequence for three different temperatures, at 0$\degree$C, 20$\degree$C and 60$\degree$C. The type of each nucleotide is represented by a colour scheme: A (white), T (Cyan), G (blue) and C (red).}
\end{center}
\end{figure}

A fraction of complementary nucleotides (A-T or G-C) on the single-stranded $\lambda$-DNA chain are close enough to form hydrogen bonds. Due to the cooperativity of base pairing, long stems with many proximal base pairs tend to form between regions of high complementarity - these are the characteristic hairpins in Fig.\ref{snapshots_ldna}. This transition between a flexible ssDNA  and significantly more rigid hairpins of dsDNA (the persistence length of dsDNA is 50 \,nm, around thirty times larger than that of ssDNA) is mediated by e.g. changes in the temperature, salt concentration or pH value. In Fig.\ref{rg_contacts_ldna} we show the radius of gyration $R_g$ and the contact fraction (the number of contacts $N_{\rm c}$ normalised by half the number of nucleotides in the ssDNA strand, which is the maximum number of possible base pairs) for the single-stranded $\lambda$-DNA versus temperature. 
\com{A contact was defined when the hydrogen-bonding interaction sites of any two nucleotides were less than 0.45 length units apart, regardless of the individual bases. In principle, this criterium cannot prevent stacked, nearest-neighbour nucleotides from being counted as a contact. Nevertheless it proved sufficiently accurate for a perfect dsDNA duplex where the number of contacts $N_{\rm c}=N/2$. Additionally, this definition will tend to include mismatched base pairs in a duplex as contacts. It will thus overestimate the number of correctly-formed Watson-Crick base pairs, but for our purposes it is more important that $2N_c$ provides a good estimate of the number of bases incorporated into hairpin structures.}

At \com{0$\degree$C, around $44\%$} of the nucleotides are involved in contacts.  When the temperature increases, the system destabilises and the number of contacts decreases significantly until it flattens out at 50$\degree$C (the same temperature at which $R_g$ has a plateau). However, while the contact fraction changes dramatically (more than a \com{factor 40 from about 0.44 to 0.01}) in this temperature range, there is only a small change in the radius of gyration (\com{around a factor 1.45 from 11.3\,nm to 16.4\,nm}).
Related snapshots from simulations at selected temperatures are given in Fig. \ref{snapshots_ldna}.

\begin{figure}[htpb]
\begin{center}
\includegraphics[width=0.5\textwidth]{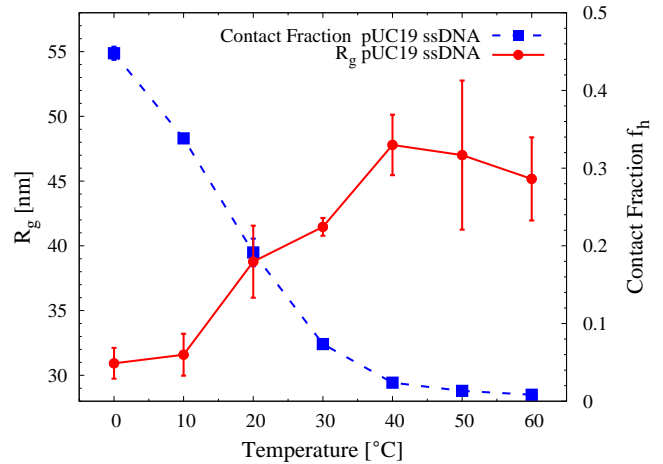}
\caption{\label{rg_contacts_puc19} Dependence of the radius of gyration $R_g$ and the fraction of direct intra-chain nucleotide contacts in the ssDNA pUC19 plasmid sequence on the temperature.
}
\end{center}
\end{figure}

\begin{figure}[htpb]
\begin{center}
\includegraphics[width=0.5\textwidth]{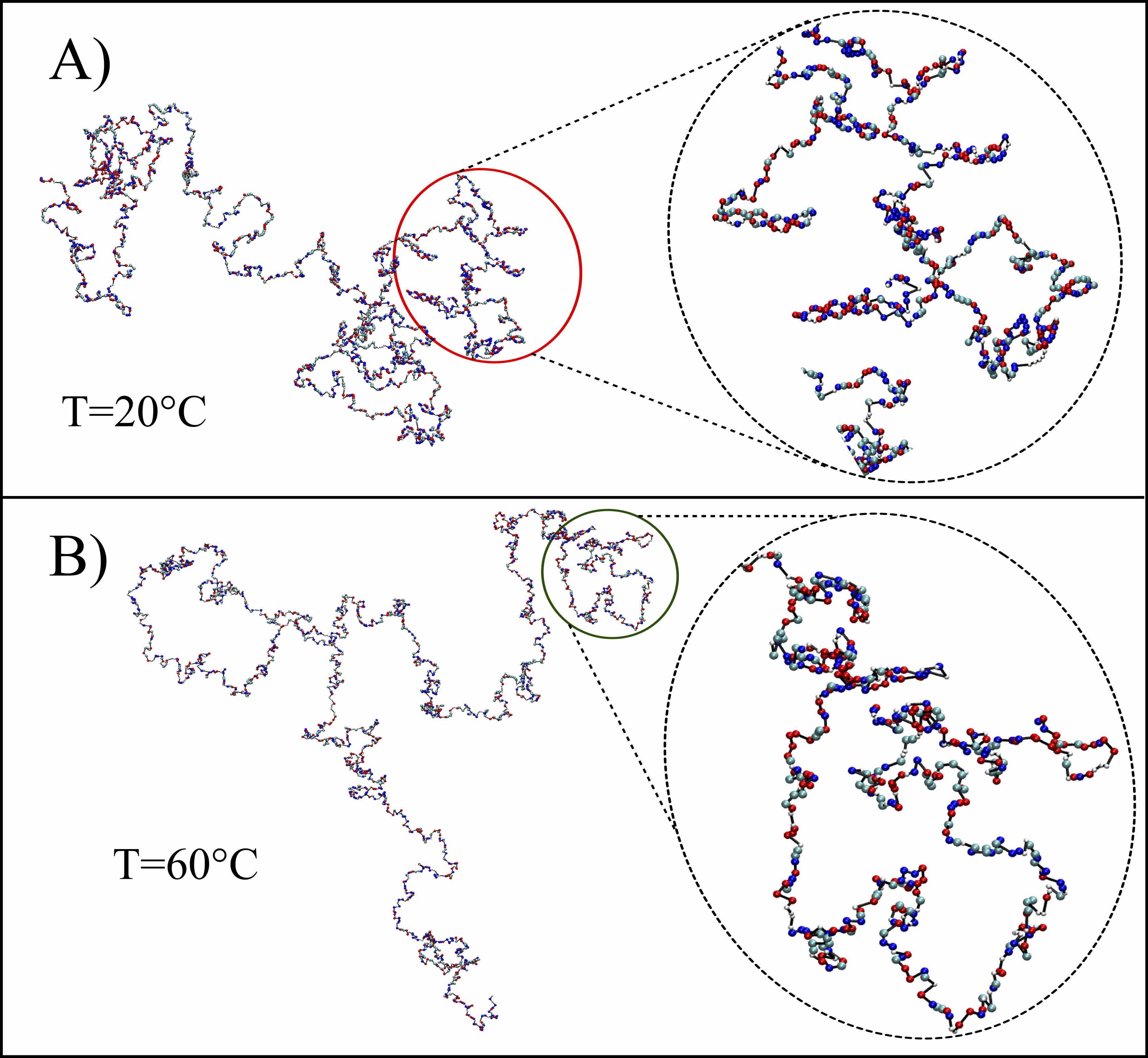}
\caption{\label{snapshots_puc19} Simulation snapshots of the pUC19-ssDNA sequence for two different temperatures, at 20$\degree$C and 60$\degree$C. The type of each nucleotide is represented by a colour scheme: A (white), T (Cyan), G (blue) and C (red).}
\end{center}
\end{figure}

We apply the same protocol as before to the pUC19 plasmid, consisting of a ssDNA sequence of 2686 nuc\-leo\-tides. For simplicity we opted for a linear molecule with freely rotating ends. The radius of gyration as a function of temperature is shown in Fig.\ref{rg_contacts_puc19}. The behaviour is very similar to the one of $\lambda$-ssDNA, particularly the minor effect that temperature changes have on $R_g$ despite dramatic changes in the number of contacts between nucleotides. While for $\lambda$-ssDNA the radius of gyration at 20$\degree$C equals 4.5\% of its total contour length, in the case of the plasmid $R_g$ represents only 2.2\%. \com{Using the theoretical expression for $R_g$ in Eq.~\ref{eq-rgnumnuc} below and monomer length $a=$0.65\,nm, Kuhn segment length $b=$2\,nm and Flory exponent $\nu=0.588$, this gives $R_g / a N = 5\%$ ($\lambda$-DNA) and $2.5\%$ (plasmid), respectively. Hence, the computational values are about 10\% smaller than the theoretical values, but generally consistent with the latter.} 
At around 50$\degree$C $R_g$ reaches a plateau, which is at least constant within the error bars. It is interesting to see that the $\lambda$-ssDNA exhibits the same tendency at the same temperature.  
As reference we also modelled the double-stranded linear pUC19 plasmid, for which we measured values of $R_g$ in the region of 130\,nm at 20$\degree$C and 170\,nm at 60$\degree$C, respectively, so about a factor 3 to 4 larger than the values of $R_g$ we obtained for the ssDNA sequence. 

In Fig.\ref{snapshots_puc19} we can see that at 20$\degree$C several nucleotides have hybridised, forming hairpin structures of 20-30\,bp located along the plasmid. When we increase the temperature of the system up to 60$\degree$C the hairpins disappear as self-hybridisation is suppressed, accounting for the substantial reduction of intra-chain contacts.

The interpretation of these results is not entirely uncomplicated as several interlinked mechanisms are at work that
all influence the radius of gyration.
When hairpins (or indeed any contact between bases) form, the hydrogen-bonding between nucleotides short-circuits all bases that are part of the hairpin, effectively shortening the contour length of the biopolymer. Hence, self-hybridisation leads to a smaller radius of gyration through a reduction of the effective contour length. Thus the smaller radius of gyration at lower temperatures can be partly explained with basic polymer physics.
On the other hand, the contribution of hairpins to the total value of $R_g$ is not zero, bearing in mind that even a rigid rod has a finite radius of gyration. The impact of self-hybridisation is thus {\it a priori} not easily assessed. Moreover, regions cut out in this way are generally bulky, tending to swell the DNA strand relative to a shorter polymer with no base pairing. This constitutes an excluded volume effect which increases $R_g$.
The exact number of hairpins and the degree of self-hybridisation depend ultimately on sequence of the ssDNA strand and are generally not quantifiable on the sole basis of polymer physics.

Nevertheless, some of the dependence of the radius of gyration on the number of formed base pairs can be rationalised using a simple and idealised physical polymer model. We assume that all nucleotide contacts are contained in well-defined hairpins. The single stranded DNA can thus be modelled as a self-avoiding polymer with attached rigid, rod-like hairpins that are cut out of the contour length of the polymer.

At high temperature the base pairing can be neglected and the genome can be modelled as a self-avoiding walk (SAW) polymer with radius of gyration

\begin{equation}
R_g= \frac{b}{\sqrt{6}} N_{\rm Kuhn}^{\nu}
\label{eq-rmsx}
\end{equation}  

\noindent where $b$ is the Kuhn segment of the polymer. At salt conditions used in the oxDNA simulations, $c_{Na}=0.2$M,  the Kuhn segment length is \com{ $b\approx$ 2\,nm~\cite{Toan2016,Sim2012,Chen2012}}.  $\nu$ is the scaling exponent~\cite{deGennes_book79} and  $N_{\rm Kuhn} = Na / b$ the number of Kuhn segments in the polymer, with $a=0.65$\,nm \cite{Toan2016,Sim2012}. Scaling exponent of a SAW polymer is $\nu=0.588$ which holds for poly-T ssDNA at physiological salt concentration~\cite{Sim2012}. Therefore, the radius of gyration is 

\begin{equation}
R_g = \frac{b}{\sqrt{6}} \left(\frac{a N}{b} \right)^{\nu}
\label{eq-rgnumnuc}
\end{equation}  

\noindent with $N$ the number of nucleotides. \com{For the $\lambda$-ssDNA sequence and the linear pUC19 plasmid this leads to 
$R_g(N=500)=16.3$\,nm and $R_g(N=2686)=43.8$\,nm, respectively.}

Assuming that all nucleotide contacts occur in hairpins, and that $2N_c$ gives a good estimate of the total number of bases cut out of the contour length by hybridisation, the effective contour length of ssDNA is reduced to $N_{\rm ss} = N - 2\,N_{\rm c}$. Consequently, the effective radius of gyration of the ssDNA is reduced to 

\begin{equation}
R_{g,ss} = \frac{b}{\sqrt{6}} \left(\frac{a (N - 2N_{\rm c})}{b} \right)^{\nu} 
\label{eq-Rgss}
\end{equation} 

\noindent depending on the number of contacts $N_{\rm c}$.
Hairpins, however, also contribute to $R_{g}$. Assuming that a hairpin is a rigid rod with length $l$ (justifiable for hairpins shorter than about 100\,nm) the radius of gyration of every hairpin is $R_{g,h} = l/\sqrt{12}$. If $k$ hairpins of equal length are formed, each hairpin will contribute

\begin{equation}
R_{g,h} = a_{ds}\,N_{\rm c} / (k \sqrt{12}) 
\label{eq-Rgh}
\end{equation}

\noindent with the effective monomer length reduced due to helicity of double stranded DNA $a_{ds} =$ 0.34\,nm. This conditions applies as all hairpins combined need to add up to the length along the contour that is in contact. 

\com{The total radius of gyration of an object is a sum over its subparts, where each subpart contributes its own radius of gyration plus a centre-of-mass distance squared, weighted by the mass.
The centre-of-mass of the total ssDNA and hairpin system is therefore }

\begin{equation}
c_m  = \frac{f_{h}}{k} \sum_{i=1}^k x_i + \frac{l}{2}\hat{n}_i
\label{eq-cm}
\end{equation}

\noindent \com{with  $x_i$ the (vector) position of the $i$-th hairpin base, i.e. the end where the hairpin is attached to the polymer. The centre-of-mass position of the $i$-th hairpin is $x_i+\frac{l}{2}\hat{n}_i$ with $\hat{n}_i$ the unit vector specifying the orientation of the hairpin's major axis. Note that only hairpins contribute because we chose the centre-of-mass of the ssDNA polymer as the origin of our coordinate system.  
The weight factor $f_{h}=2N_{\rm c}/N$ is determined by the fraction of total polymer mass contained in the hairpins. The quantity $f_h$ is equal to the contact fraction shown in Figs.~\ref{rg_contacts_ldna} and~\ref{rg_contacts_puc19}.
The total radius of gyration of the ssDNA and hairpins system becomes }
 
\begin{equation}
R_g^2 = (1-f_h) (R_{g,ss}^2 + c_m^2) + \frac{f_h}{k} \sum_{i=1}^k  R_{g,h}^2 + (x_i +\frac{l}{2}\hat{n}_i - c_m)^2
\label{eq-Rgtot}
\end{equation}  

\noindent \com{where the first term on the right hand side is the contribution of the ssDNA and the second term, the sum, is performed over all $k$ hairpins. Note that the fraction of total mass in each hairpin is $f_h/k$ and $x_i +\frac{l}{2}\hat{n}_i - c_m$ is the distance between the hairpin centre-of-mass and single-stranded polymer centre-of-mass. }

\com{Assuming that the positions of hairpins are uniformly random and uncorrelated, inserting Eq.~(\ref{eq-cm}) into Eq.~(\ref{eq-Rgtot}) and employing some basic algebra outlined in Appendix \ref{theory},  the expected value for the squared radius of gyration is obtained }

\begin{equation}
\langle R_g^2 \rangle = R_{g,ss}^2  \left( 1- \frac{f_h^2}{k}\right) + R_{g,h}^2\left( 4 f_h-3\frac{f_h^2}{k}\right) 
\label{eq-avRgtot}
\end{equation}  

\noindent \com{with $R_{g,ss}$ and $R_{g,h}$ given by Eqs.~(\ref{eq-Rgss}) and (\ref{eq-Rgh}), respectively, and the contact fraction $f_h=2N_{\rm c} / N$. }

\begin{figure}[htbp]
\begin{center}
\includegraphics[width=0.5\textwidth]{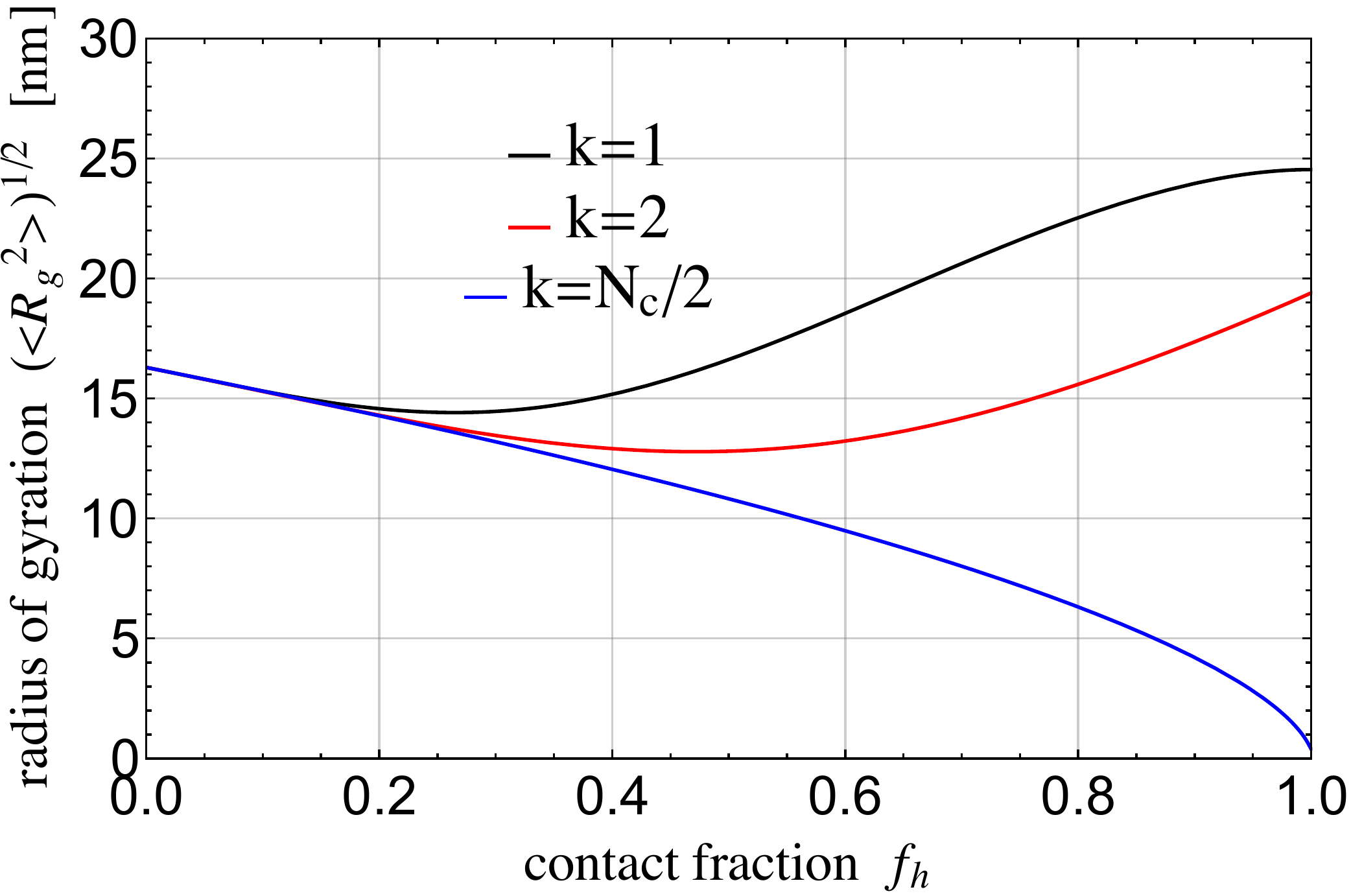}
\caption{\label{fig-thRg_contacts} Radius of gyration $\sqrt{\langle R^2_g \rangle}$ as a function of the contact fraction $f_{h}=2N_{\rm c} / N$ for different number of formed hairpins $k$. The curves were obtained from Eq.~\ref{eq-avRgtot} using the following parameters: \com{Kuhn segment $b$=2\,nm}, nucleotide size $a$=0.65\,nm, double stranded nucleotide effective size $a_{ds}=$0.34\,nm, scaling exponent $\nu=0.588$, number of nucleotides $N=500$.}
\end{center}
\end{figure}

\com{We have assumed that hairpins do not interact with the ssDNA polymer, or with other hairpins, and that all $k$ hairpins are of the same length.} However, even this relatively simple, idealised derivation de\-mon\-strates that the radius of gyration depends on both the number of contacts and the hairpin length. This is shown in Fig.~\ref{fig-thRg_contacts} for a sequence of $N=500$ nucleotides, i.e. the length of our $\lambda$-ssDNA. The dependence on the number of contacts is obviously non-monotonous. The values $k=1$ and $k=N_{\rm c}/2$ (assuming a hairpin needs at least 2 contacts to be labelled as a hairpin) are the limits of the possible hairpin distribution and corresponding values for $\langle R^2_g \rangle$ provide the upper and lower physical limit for the expected value of the radius of gyration. 
\com{The simulations, Fig.~\ref{rg_contacts_ldna}, result in a radius of gyration around 12.6\,nm and  11.3\,nm at the observed contact fraction of around 32\% and  45\%, respectively, in good agreement with the theoretical prediction shown on Fig.~\ref{fig-thRg_contacts}}. We also see that for an even larger number of contacts the possible range of values of $\langle R^2_g \rangle$ is quite wide. This is of course a much idealised and simplified reasoning, but it elucidates the non-trivial nature of these interdependencies.\com{The theory neglects the excluded volume of regions cut out of the contour length by hybridisation; taking this into would increase the $R^2_g$ in Eq.~(\ref{eq-Rgtot}), while additional bases cut out by hybridisation but not contributing to $N_c$ would decrease it. We speculate that the two effects cancel out, to a degree, resulting in a good agreement between theory and simulations.}

%%%%%%%%%%%%%%%%%%%%%%%%%%%%%%%%%%%%%%%%%%%%%%%%%
% Conclusions
%%%%%%%%%%%%%%%%%%%%%%%%%%%%%%%%%%%%%%%%%%%%%%%%%

\section{Conclusions}\label{conclusions}

The implementation of the oxDNA model for coarse-grained DNA modelling into a community molecular dynamics code such as LAMMPS 
reduces the entry barrier of using the model significantly. Moreover, it allows to combine this coarse-grained
force field with different features that are already enabled in LAMMPS. 

The Langevin-type rigid-body integrators that are distributed together with the LAMMPS USER-package, particularly the DOT-C integrator, 
offer additional advantages over the existing standard rigid-body integrators for Langevin dynamics. 
They show improved stability at the costs of a very small overhead. This permits larger timesteps and therefore larger  
physical simulation times.

The parallel performance of the MPI-only implementation, as demonstrated through scaling tests using a simple benchmark,
 is excellent provided there are at least a few dozen particles per MPI-task.  These results show effectively that the oxDNA 
model is well suited for large and extremely large problems in DNA and RNA modelling. It can tackle problem sizes that 
were well beyond the reach of the original standalone implementation of the model. 
It is worth mentioning that the GPU-accelerated version of the standalone code is also limited to speedups of typically 
a factor 30 compared to the single core performance. 
Based on the scaling analysis of the benchmarks it could be said that this is matched by the performance of a single
multi- or many core chip.

The applications we opted for, a sequence of linear, single-stranded $\lambda$-bacteriophage and pUC19 plasmid DNA, 
are motivated primarily by currently ongoing projects in the under-investigated area
of single-stranded DNA, rather than by an attempt to harvest the performance of the new \\LAMMPS implementation. 
The results shows that the conformation of ssDNA is strongly affected by the tendency 
to self-hybridise upon cooling, i.e. to form intra-chain base pairs between complementary nucleotides 
on the same strand that lead to hairpins, local regions of dsDNA, and less structured domains 
of clustered nucleotides.
The radius of gyration $R_g$ of both ssDNA examples is predicted to be relatively insensitive towards
temperature changes between \com{0$\degree$C and 60$\degree$C}. The slight reduction of $R_g$ can be at least 
partly explained with a shorter effective contour length of the biopolymer due to hairpin formation.
This explanation, however, disregards some of the more subtle intricacies of the self-hybridisation process.
Hairpins contribute as well to the total value of $R_g$. The hybridised domains of clustered nucleotides introduce  
an excluded volume effect, which increases the radius of gyration. Last but not least, the DNA sequence 
determines whether any self-hybridisation can occur in the first place.
\com{It should be noted that there is a large number of possible self-hybridised bonding configurations.
This means that the system is likely to fall into a particular one upon quenching and to remain there. 
However, by using a number of independent configurations we have presumably
reached states that are representative, although these are not guaranteed to be the most stable ones.}

In the future it may be possible to focus on ring mo\-le\-cules that contain superhelical twist and have different 
number of helical turns compared to their natural form. These rings may be opened by introducing a single strand break,
which releases the superhelical twist, a mechanism that is known to be highly relevant during gene replication and expression.

\begin{acknowledgements}
This work was funded under the embedded CSE programme of the 
ARCHER UK National Supercomputing Service (eCSE05-10).
YAGF acknowledges support from the Mexican National Council for Science and Technology
(CONACyT, PhD Grant 384582).
TC acknowledges support from the Herchel Smith Scholarship and the CAS PIFI Fellowship.
TEO acknowledges his Royal Society University Research Fellowship.
OH acknowledges support from the EPSRC Early Career Fellowship Scheme
(EP/N019180/2). 
\end{acknowledgements}

\section*{Authors contributions}
OH, YAGF, TC and TEO designed and performed research, analysed data, and wrote the paper. 
The implementation was undertaken in a collaboration between OH and TEO.
%
% BibTeX users please use
%\bibliographystyle{epj}
\bibliography{si_epje}
%

%%%%%%%%%%%%%%%%%%%%%%%%%%%%%%%%%%%%%%%%%%%%%%%%%
% Appendix
%%%%%%%%%%%%%%%%%%%%%%%%%%%%%%%%%%%%%%%%%%%%%%%%%

\appendix
\section*{Appendix}

\section{Profiling}\label{profiling}

\begin{figure*}[htbp]
\centering
\begin{minipage}{0.92\textwidth}
\centering
\includegraphics[width=0.44\textwidth]{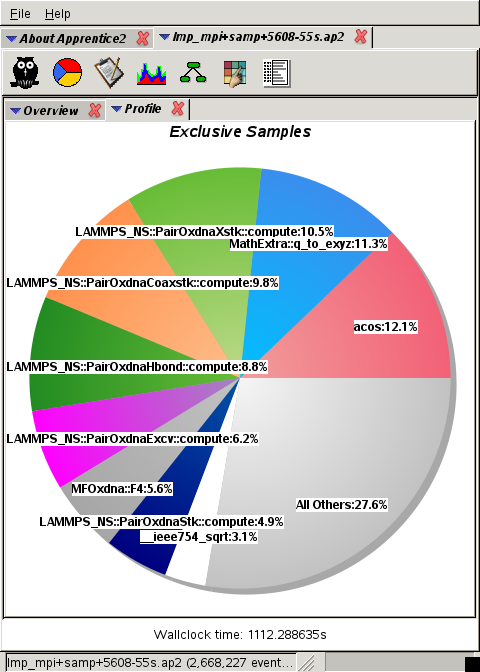}
\includegraphics[width=0.443\textwidth]{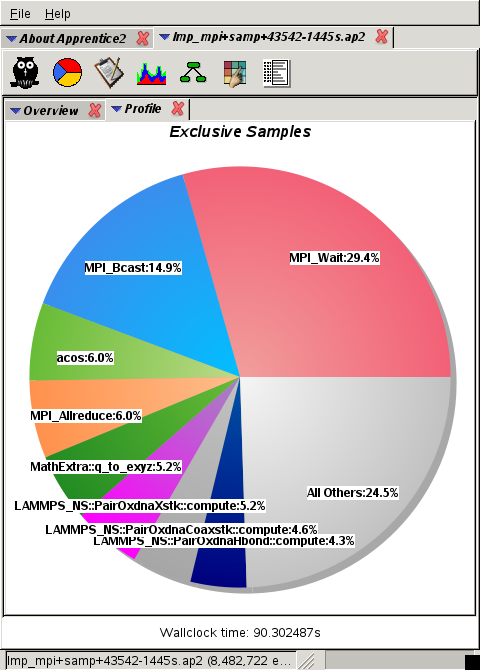}
\caption{\label{perf-60kbps} Craypat performance analysis of a sampling experiment for the low density benchmark (60 kbp) on a single node (left, 24 MPI-tasks) and for 2048 MPI-tasks (right). Note that the assigned colour code for the functions is different in both cases.}
\end{minipage}
\end{figure*}
\begin{figure*}[htbp]
\centering
\begin{minipage}{0.92\textwidth}
\centering
\includegraphics[width=0.43\textwidth]{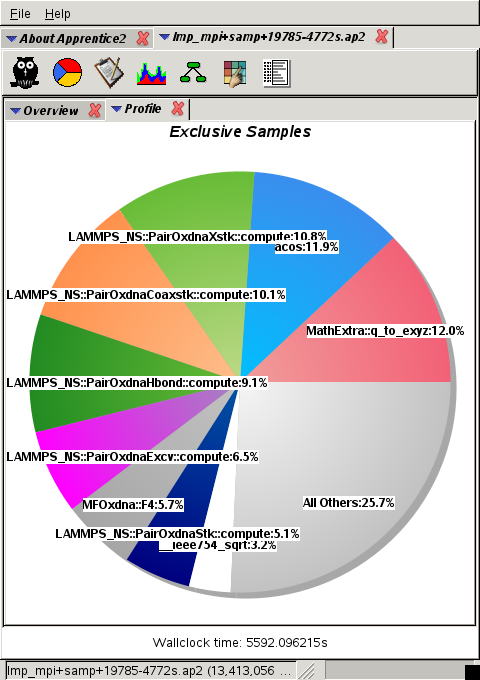}
\includegraphics[width=0.437\textwidth]{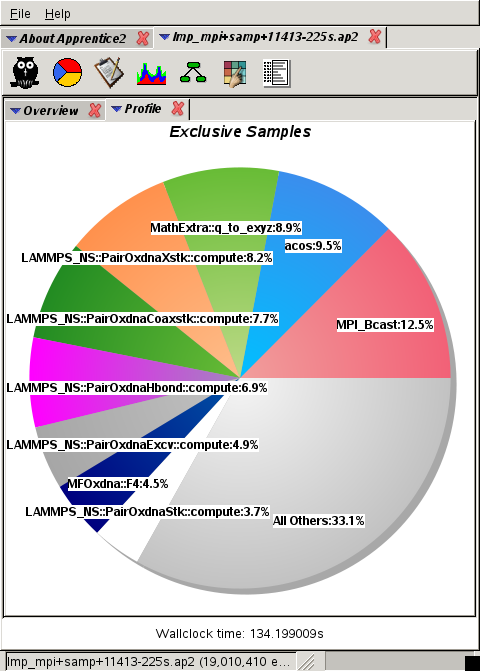}
\caption{\label{perf-1mbps}Craypat performance analysis of a sampling experiment for the high density benchmark (960 kbp) on a single node (left, 24 MPI-tasks) and for 2048 MPI-tasks (right). Note that the assigned colour code for the functions is different in both cases.}
\end{minipage}
\end{figure*}

Profiling allows a detailed analysis of the implementation and gives an overview of how much time the code spends in each 
individual subroutine. 
We used the Craypat Performance Tools on the ARCHER UK National Supercomputing Service to conduct sampling experiments of the 
high and low density benchmarks.
Although the experiments where actually performed with the oxDNA model, the results are representative as well for oxDNA2 as the 
only difference between the two is a different local geometry of the interaction sites and an additional pair interaction in
form of a Debye-H\"uckel potential.

Fig. \ref{perf-60kbps} shows a pie chart
of the low density (LD) run. The image on the left shows the results on a single node with 24 MPI-tasks, whereas the image on the right 
is for 2048 MPI-tasks.
Focussing first on a single node, calls to the MPI-library are below $5\%$ and do not appear with an individual pie section. 
The total time spent in the force calculation is around $86\%$ (according to the LAMMPS breakdown).
Interestingly, a significant fraction of the time is spent on calculating the local body coordinate system of the nucleotide from the quaternion 
degrees of freedom (MathExtra::q\_to\_exy, $11.3\%$).

A significant portion falls also on the calculation of the inverse cosine (acos, $12.1\%$). 
The conversion from quaternions to 3-vectors is done separately in every single interaction. 
This has been done for simplicity, but represents a 6-fold overhead as it could be optimised by calculating the 3-vectors only once per timestep,
then saving the for later use by the interactions. This optimisation would come at increased communication as the additional nine
components of the three unit vectors would have to be communicated across the process boundaries.
Another possibility, and a major adaptation, would be to formulate the entire force calculation in generalised quaternion forces and torques, 
therefore avoiding the transformation in the first place. We decided deliberately against this possibility as this would require 
calculation of four force and torque components in quaternion space. The calculation with 3-vectors on the other hand, as currently implemented, 
requires only three force and torque components. Perhaps most importantly, they can be made available directly to the other LAMMPS routines. 
It is thus very likely that a performance gain from avoiding the transformation would be outweighed either by the larger number of 
additional components and generalised quaternion forces and torques which also had to be communicated across the process boundaries 
or by disadvantages from a software engineering point of view.

The large fraction of the inverse cosine is more difficult to optimise. It emerges in the stacking, cross- and coaxial stacking
and hydrogen bonding interactions through a partial derivative with respect to the relative distances. A previous version of the 
implementation spent a whopping $29\%$ of its time calculating the inverse cosine. This prohibitively large figure could be 
cut down to the current $12 \%$ by introducing appropriate early-rejection criteria in each force calculation. Further improvements
might be possible through small-argument approximations of the inverse cosine. This will be tested in a future version
of the code (e.g. for the upgrade to oxDNA 2.0).

At 2048 MPI-tasks, shown on the right of Fig. \ref{perf-60kbps}, the code spends more than $50\%$ of its time in call to the MPI-library.
The percentage of time in the force calculation has fallen to about $43\%$. As stated above, this is primarily the consequence of 
an insufficient number of local atoms with respect to the number of ghost atoms, and does not reflect a problem with the 
parallel performance of the implementation.  

For the HD benchmark on a single node, shown on the left in Fig. \ref{perf-1mbps}, calls to the MPI-library are below $3\%$.
The conversion of quaternions to 3-vectors \\
(MathExtra::q\_to\_exyz) and the calculation of the inverse cosine (acos) are constant 
at about $12\%$. At 2048 MPI-tasks we observe a parallel efficiency of about $85\%$. The time spent in the force calculation is
still about $82\%$ (according to the LAMMPS breakdown) with calls to the MPI-library amounting to just below $13\%$. 
The CPU time of the quaternion conversion to the local body frame of the nucleotide and the inverse cosine each at are around $9\%$ 
due to the larger share of the calls to the MPI-library.

\section{Derivation of \boldmath{$\langle R^2_g \rangle$}}\label{theory}

We assume that the position of hairpin bases, as well as the orientation of hairpins, is uniformly  random and uncorrelated along the ssDNA contour, formally: $\langle x_i \rangle=0$, $\langle x_i^2 \rangle = R_{g,ss}^2$, $\langle x_i x_j \rangle = 0$ for $i\ne j$, and similarly for the orientation: $\langle \hat{n}_i \rangle =0$, $\langle \hat{n}_i^2 \rangle =1$, $\langle \hat{n}_i \hat{n}_j \rangle =0$ for $i\ne j$,  $\langle x_i \hat{n}_j \rangle =0$. 
These properties result in $$\langle c_m \rangle = 0$$ and $$\langle c_m^2 \rangle = \frac{f_h^2}{k} (R_{g,ss}^2 + l^2/4).$$

\noindent The average $R_g^2$ becomes  
\begin{eqnarray}
\langle R_g^2 \rangle &=& (1-f_h) R_{g,ss}^2 + (1-f_h) \langle c_m^2 \rangle + f_h R_{g,h}^2 + \nonumber \\ 
&+& \frac{f_h}{k}\Big\langle \sum_i (x_i +\frac{l}{2}\hat{n}_i - c_m)^2 \Big\rangle \;.
\end{eqnarray}

\noindent The average of the sum is 
\begin{eqnarray}
&&\Big\langle \sum_{i=1}^k (x_i + \frac{l}{2}\hat{n}_i - c_m)^2 \Big\rangle = \sum_{i=1}^k  \Big\{\langle x_i^2 \rangle + \langle c_m^2 \rangle + \frac{l^2}{4} \langle \hat{n}_i^2 \rangle +\nonumber\\
&& \hspace*{1cm} +\;  l \langle x_i \hat{n}_i \rangle - 2 \langle x_i c_m \rangle - l \langle \hat{n}_i c_m \rangle\Big\}   \nonumber \\
&& = k R_{g,ss} + f_h^2  (R_{g,ss}^2 + l^2/4) +  k \frac{l^2}{4} - 2f_h R_{g,ss}^2 - f_h \frac{l^2}{2}\nonumber\\  
\end{eqnarray}

\noindent using that $\sum_i \langle x_i c_m \rangle = \frac{f_h}{k}\sum_{ij} \langle x_i (x_j + \frac{l}{2} \hat{n}_j)\rangle = f_h R_{g,ss}^2$  and  $\sum_i \langle \hat{n}_i c_m \rangle =  f_h \frac{l}{2}$. Furthermore, $l^2=12R_{g,h}^2$.

Using these relations the expected value for the squared radius of gyration is obtained

\begin{equation}
\langle R_g^2 \rangle = R_{g,ss}^2  \left( 1- \frac{f_h^2}{k}\right) + R_{g,h}^2\left( 4 f_h-3\frac{f_h^2}{k}\right)\;, 
\end{equation}  

\noindent with $R_{g,ss}$ and $R_{g,h}$ given by Eqs.~(\ref{eq-Rgss}) and (\ref{eq-Rgh}), respectively, and the contact fraction $f_h=2N_{\rm c} / N$.

\end{document}